\documentclass[]{aa}

\usepackage{longtable}
\usepackage{threeparttablex}
\usepackage{graphicx}
\usepackage{supertabular}

\usepackage{txfonts}
\newcommand{\kms}{\,km\,s$^{-1}$}

\newcommand{\vorid}{V1192\,Ori}
\newcommand{\vori}{V1192\,Ori }

\begin{document}

   \title{Antisolar differential rotation with surface lithium enrichment on the single K-giant V1192\,Ori \thanks{Based on data obtained with the STELLA robotic observatory in Tenerife, an AIP facility jointly operated by AIP and IAC.}}

\author{Zs.~K\H{o}v\'ari\inst{1}
\and K.~G. Strassmeier\inst{2}
\and T.~A. Carroll\inst{2}
\and K.~Ol\'ah\inst{1}
\and L.~Kriskovics\inst{1}
\and E.~K\H{o}v\'ari\inst{3}
\and O.~Kov\'acs\inst{1,3}
\and K.~Vida\inst{1}
\and T.~Granzer\inst{2}
\and M.~Weber\inst{2}
}

\offprints{Zs. K\H{o}v\'ari}

\institute{Konkoly Observatory,
Research Centre for Astronomy and Earth Sciences, Hungarian Academy of Sciences,
Konkoly Thege \'ut 15-17., H-1121, Budapest, Hungary\\
  \email{kovari@konkoly.hu}
  \and Leibniz-Institute for Astrophysics Potsdam (AIP), An der Sternwarte 16,
D-14482 Potsdam, Germany
\and E\"otv\"os University, Department of Astronomy, Pf. 32., H-1518, Budapest, Hungary}

   \date{Received ; accepted}

\abstract  
{Stars with about 1$-$2 solar masses at the red giant branch (RGB) represent an intriguing period of stellar evolution, i.e. when the convective envelope interacts with the fast-rotating core. During these mixing episodes freshly synthesized lithium can come up to the stellar surface along with high angular momentum material. This high angular momentum may alter the surface rotation pattern.}
{The single rapidly rotating K-giant \vori is revisited to determine its surface differential rotation, lithium abundance, and basic stellar properties such as a precise rotation period. The aim is to independently verify the antisolar differential rotation of the star and possibly find a connection to the surface lithium abundance.}
{We applied time-series Doppler imaging to a new multi-epoch data set. Altogether we reconstructed 11 Doppler images from spectroscopic data collected with the STELLA robotic telescope between 2007--2016. We used our inversion code  \emph{iMap} to reconstruct all stellar surface maps. We extracted the differential rotation from these images by tracing systematic spot migration as a function of stellar latitude from consecutive image cross-correlations.}
{The position of \vori in the Hertzsprung-Russell diagram suggests that the star is in the helium core-burning phase just leaving the RGB bump. We measure $A({\rm Li})_{\rm NLTE}=1.27$, i.e. a value close to the anticipated transition value of 1.5 from Li-normal to Li-rich giants. Doppler images reveal extended dark areas arranged quasi-evenly along an equatorial belt. No cool polar spot is found during the investigated epoch. Spot displacements clearly suggest antisolar surface differential rotation with $\alpha=-0.11\pm0.02$ shear coefficient. }
{The surface Li enrichment and the peculiar surface rotation pattern may indicate a common origin. }


\authorrunning{K\H{o}v\'ari et al.}
\titlerunning{Antisolar differential rotation and surface lithium enrichment on V1192\,Ori}

\maketitle

%

\section{Introduction}

Stellar magnetic activity manifests itself in cool starspots on the stellar surface and is strongly related to rapid rotation. Although most of the stellar angular momentum is supposed to be transferred to its environment by a wind and consequent magnetic braking on the main sequence, there are examples of evolved red giant branch (RGB) stars that are still rapidly rotating and magnetically active. Most of these are members of close binary systems in which tidal forces maintain fast rotation. Rapidly rotating single giants remain a challenge for angular-momentum evolution theories. There are scenarios that allow a star to keep its angular momentum after the main sequence, such as enhanced mixing and dredge-up episodes \citep{1989ApJ...346..303S,2010A&A...522A..10C} or planet engulfment \citep{2016A&A...593A.128P}.

After the main sequence, RGB stars of about 1$-$2 solar masses represent an intriguing episode of stellar evolution. When the deepening convective envelope interacts with the fast-rotating core, angular momentum is transported to the surface. The mixing of the envelope material with the hotter layers below infer the decay of light elements. This short evolutionary period called the first dredge up is responsible for the dilution of the surface lithium. Despite the expected low lithium abundance, a handful of low-mass stars in this evolutionary state show lithium enrichment on their surfaces. For intermediate mass ($\approx$4$-$7\,$M_{\odot}$) stars at the asymptotic giant branch (AGB), the Cameron--Fowler mechanism \citep{1971ApJ...164..111C} followed by a transport of the Li to cooler regions can plausibly explain the lithium excess. On the other hand, for lower masses at the RGB this mechanism may not work and other non-standard extra mixing processes are required \citep[cf.][]{1992ApJ...392L..71S,2010A&A...522A..10C}. In turn, cool-bottom processes \citep{1995ApJ...447L..37W,1999ApJ...510..217S} below the convection zone may also be capable for transporting material to layers hot enough for the Cameron--Fowler mechanism and then returning to the convection zone. An alternative evolutionary model was proposed by \citet{1996ApJ...456L.115D} in which basically every low-mass K-giant undergoes a short ($\approx$10$^5$yr) Li-rich phase on the RGB \citep[see also][]{1997ApJ...482L..77D,2016ApJ...819..135K}. Extra non-axisymmetric mixing that leads to an inhomogeneous super-granulation pattern on the surface in the form of large cool and warm features was invoked to explain super-meteoritic Li abundances in \citet{2015A&A...574A..31S}.

So far, differential rotation is found to be solar type for main-sequence stars, but giant stars can exhibit antisolar differential rotation, i.e. the equator rotates more slowly than the poles  \citep[][etc.]{1991LNP...380..297V,2003A&A...408.1103S,2015A&A...573A..98K}.
Theoretically, this phenomenon is induced and maintained by strong meridional circulation \citep{2004AN....325..496K}. \citet{2007Icar..190..110A} proposed
that strong mixing by turbulent convection would be the primary agent for angular momentum equilibration and thus antisolar differential rotation. Detailed numerical simulations by, for example \citet{2014A&A...570A..43K} and \citet{2014MNRAS.438L..76G}, suggest that rapidly rotating stars with a small Rossby number yield solar-like differential rotation, while weakly rotating stars with large Rossby numbers may sustain antisolar differential rotation. On the other hand, only three single rapidly rotating giants are known so far to exhibit antisolar differential rotation, namely DI\,Psc, DP\,CVn \citep{2013A&A...551A...2K}, and \vori \citep[][the star revisited in this paper]{2003A&A...408.1103S}.

Interestingly, these stars are listed among the very few fast-rotating RGB stars with unusually high surface lithium abundances \citep[see Table~1 in][]{2000A&A...359..563C}, implying a possible connection between the antisolar differential rotation profile and the enhanced surface lithium  \citep{2014A&A...571A..74K}. We posit whether it is possible that a yet-unknown mixing mechanism responsible for the lithium enrichment can eventually also alter the surface differential rotation profile to be antisolar. We believe that such fast-rotating, single, RGB stars provide a good opportunity to investigate the relationship between activity, rotation, differential rotation, and surface lithium abundance.

In this paper, we present a time-series Doppler imaging study of the single rapidly rotating
($P_{\rm rot}\approx28$\,days) K-giant \object{V1192\,Ori} (=HD\,31993), based on spectroscopic
observations from the STELLA robotic observatory in Tenerife \citep{2010AdAst2010E..19S}.
The star was found to have strong Ca\,{\sc ii} H\&K emission and was classified as a K2 giant \citep{1973AJ.....78..687B,1986ApJS...60..551F,1990ApJS...72..191S} with an unusually high $v\sin i$ value of 31\kms\ \citep[see also][]{1997PASP..109..514F,1993ApJ...403..708F}. \vori was catalogued as Li-rich according to its lithium abundance, which is substantially larger than expected for an ordinary K giant \citep{1988ESASP.281a.331F}. The non-local thermodynamic equlibrium (NLTE) lithium abundance for \vori was measured as $A({\rm Li})=1.4\pm0.2$ \citep{1993ApJ...403..708F} on the usual logarithmic scale, i.e. just at the border of $A({\rm Li})\ge1.5$  between Li-normal and Li-rich giants \citep[cf.][]{2000A&A...359..563C}. \citet{2000A&A...364..674C} found $A({\rm Li})$ of 1.7 assuming local thermodynamic equlibrium (LTE), and 1.8 with NLTE correction. More recently, \citet{2015AJ....150..123R} quoted the star as Li-normal with $A({\rm Li})=1.26$ under NLTE assumption.

\vori shows a wide range of magnetic-activity indicators from X-ray to infrared (but remains undetected in radio; see \citealt{1987MNRAS.229..659S}). The star was listed as an X-ray source by the \emph{Einstein} and ROSAT surveys \citep[][respectively]{1990ApJS...72..567G,1999A&A...349..389V}. International Ultraviolet Explorer (IUE) observations revealed an active UV chromosphere \citep{1993ApJ...403..708F} in accordance with the photospheric light variability \citep{1997A&AS..125...11S,1999A&AS..140...29S}, which is attributed to stellar rotation and cool starspots. According to the Zeeman signatures detected by \citet{2015A&A...574A..90A}, \vori possesses a strong surface magnetic field that is likely produced by an $\alpha\Omega$-type dynamo. The star is also listed in the IRAS \citep{1999yCat.2225....0G} and 2MASS \citep{2003yCat.2246....0C} infrared point source catalogues.

The first comprehensive photometric and spectroscopic study of \vori was carried out by \citep[][hereafter Paper~I]{2003A&A...408.1103S}. A rotational period of $\approx$26 days was derived from the
photometric variability which, together with $v\sin i$ of 32\kms ,  suggested a minimum radius of $\approx$16\,$R _{\odot}$ that is consistent with the K2 giant classification. The comparison of the position of \vori in the Hertzsprung-Russell (H-R) diagram with feasible evolutionary tracks yielded a mass determination of 1.9\,$M _{\odot}$. In Paper~I two Doppler images were presented for consecutive rotation periods  showing cool starspots mostly at low to mid-latitudes. Surface differential rotation was investigated by cross-correlating the subsequent maps and yielded antisolar differential rotation. Such observations are relevant constraints for dynamo theory, hence their reliability is of great consequence. Therefore, we revisit \vori and carry out a new, more detailed Doppler-imaging study from new high-quality spectroscopic data.

The paper is organized as follows. In Sect.~\ref{obs} we describe the observations and in Sect.~\ref{astro} provide a more accurate photometric period from the available photometric data. The astrophysical data are summarized in Sect.~\ref{astrop}, where we also present a redetermination of the surface lithium abundance. In Sect.~\ref{di}, we focus on the time-series Doppler imaging. The results are summarized and discussed in Sect.~\ref{disc}.

\section{Observations}\label{obs}

\subsection{Photometry}\label{obs_phot}

\begin{figure}[]
{\bf a.}\\
\includegraphics[angle=0,width=0.974\columnwidth]{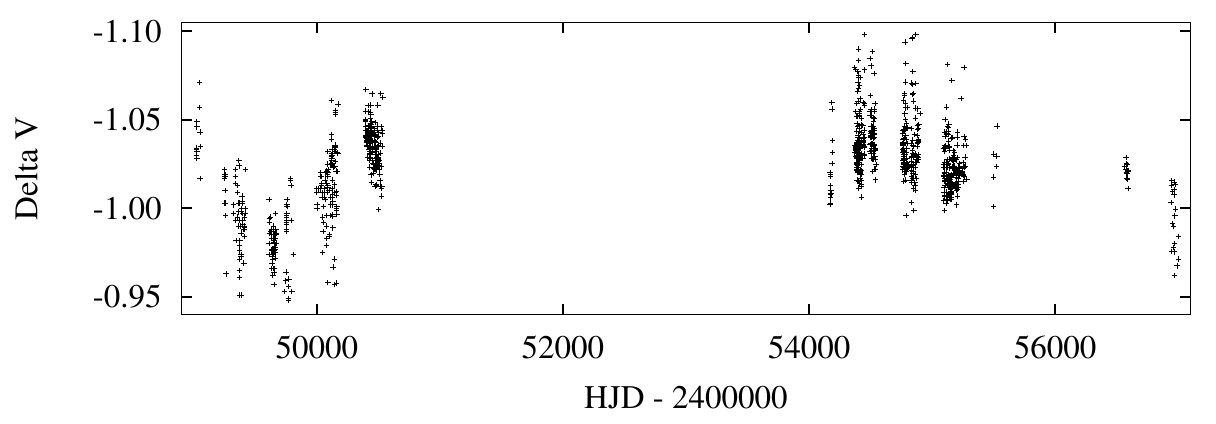}
{\bf b.}\\
\includegraphics[angle=0,width=1.0\columnwidth]{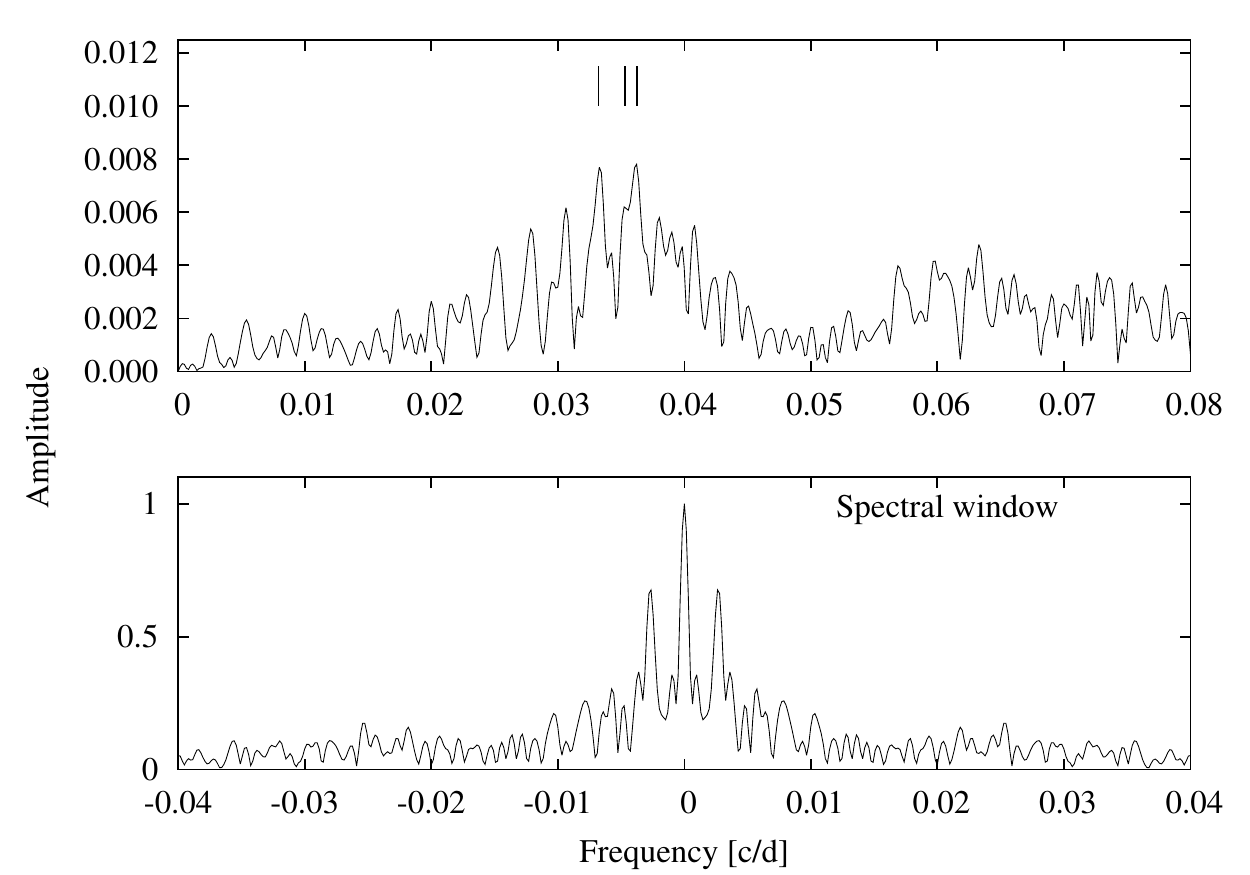}
\caption{{\bf a.} Long-term photometric $V$ data of \vori observed with the Amadeus APT. {\bf b.} Amplitude spectrum (top) and spectral window (bottom) from the Johnson $V$ data obtained between 1993--1997. The three most significant period signals are indicated. See text for details. }
\label{aptdatafft}
\end{figure}

Two data sets for \vori were obtained. The first part of the observations was collected between February 1993--March 1997 (JD\,2,449,024--2,450,537), while the second part was obtained between March 2007--December 2014 (JD\,2,454,173--2,457,003). All observations were carried out with the T7 (`Amadeus') 0.75\,m automatic photoelectric telescope (APT) at Fairborn Observatory in southern Arizona  \citep{1997PASP..109..697S}, which is currently owned and operated by the Leibniz-Institute for Astrophysics Potsdam (AIP); see \citet{2001AN....322..325G} for more details. The two data sets consist of altogether 1312 measurements in Johnson-Cousins $V$ and
$I_C$. HD\,32191 ($V=8\fm520\pm0\fm015$, $I_C=7\fm287\pm0\fm021$) was used as a comparison star while  HD\,32073 was used as the check star. Typical data quality is around 6\,mmag rms in the first data set but significantly worse in the second because of a defocussing problem. The photometric $V$ data are plotted in Fig.~\ref{aptdatafft}.

\subsection{Spectroscopy}

A total of 460 high-resolution echelle spectra were recorded with the 1.2\,m STELLA robotic observatory \citep{2010AdAst2010E..19S} at the Iza\~{n}a Observatory in Tenerife, Spain, between Jan\,9, 2007 and Feb\,3, 2016. The telescope is equipped with the fibre-fed, fixed-format STELLA Echelle Spectrograph (SES). The spectra cover the full 3900--8800\,\AA\ wavelength range with an average spectral resolution of $R=55\,000$. Further details on the performance of the system and the
data reduction procedures can be found in \citet{2008SPIE.7019E..0LW,2012SPIE.8451E..0KW} and \citet{2011A&A...531A..89W}. Table~\ref{Tab1} in the Appendix gives the log of the SES observations used for the Doppler reconstructions presented in Sect.~\ref{di}.

\section{Photometric period}\label{astro}

Using the Fourier transformation-based frequency analyser code MuFrAn \citep{2004ESASP.559..396C}
we analysed the $V$ data of \vori to refine the photometric period. In Paper~I, only the
1996-1997 APT data were used for the period search, yielding in principle the correct but ambiguous and uncertain period of 25.3\,d.

The quality of the new photometric data turned out to be very uneven between the two sets of observations (see Sect.~\ref{obs_phot}). We found that the second set from March 2007--December 2014 is rather noisy compared to the first set from February 1993--March 1997. Moreover, the rotational variation of \vori has usually low amplitude. Therefore, we used only the first data part for the period analysis. The resulting amplitude spectrum is shown in Fig.~\ref{aptdatafft}. For the refinement of the period we took the three highest peaks and perform a multi-periodic fit.  From this the highest amplitude belongs to the 28.3\,d period (middle tick mark in the top panel of Fig.~\ref{aptdatafft}). We confirmed this period by consecutively pre-whitening with the three periods. Finally, we settled on $P_{\rm phot}=28.30\pm0.02$\,d. The other significant periods denoted by the other tick marks are 27.59\,d and 30.06\,d. Owing to the crosstalk between the neighbouring peaks, it is not evident that the highest amplitude peak corresponds to the middle tick. Therefore, in Table~\ref{TFT} we listed the frequencies, amplitudes and residuals of 1-, 2-, 3- and 4-component Fourier-fits carried out subsequently. The residuals decrease significantly until the 3-component fit, however, introducing the 4th component yields only marginal improvement.
In the 3- and 4-component fits the highest amplitudes correspond to $\approx$0.0353 c/d, i.e. $\approx$28.3 days period.

\begin{table}[]
\caption{Multi-component Fourier-fit parameters} \label{TFT}
 \centering
 \begin{tabular}{cccc}
  \hline\noalign{\smallskip}
   Frequency & Amplitude & Residual & Residual \\
$$[c/d]  &    &   & difference\\
   \hline\hline
  \noalign{\smallskip}
$f_1$=0.036250  & 0.00777  & 0.012761 &     \\
  \hline
  \noalign{\smallskip}
$f_1$=0.036214  & 0.01034  &  &     \\
$f_2$= 0.032574  &  0.00819  & 0.011528 &     0.001233\\
    \hline
  \noalign{\smallskip}
 $f_1$=0.036236  & 0.00922  &  &     \\
$f_2$=0.033289  &  0.00648  &  &     \\
  $f_3$=0.035346  & 0.01090  & 0.010278 &    0.001250 \\
    \hline
  \noalign{\smallskip}
$f_1$= 0.036394  & 0.00658  &  &     \\
$f_2$=0.033197  &  0.00928  &  &     \\
$f_3$= 0.035302  & 0.00940  &  &     \\ 
$f_4$=0.030785  & 0.00556  & 0.009997 &     0.000281\\ 
\hline\hline
 \end{tabular}
\end{table}

By interpreting the three period peaks as a sign of surface differential rotation, we obtained a surface shear parameter $\Delta P/P$ of $\approx$0.09. This shear parameter is within the error box of the value derived from consecutive Doppler images in Paper~I. From photometric time series alone one cannot determine the sign of the surface shear, that is, whether the differential rotation is solar type or antisolar, but see  \citet{2015A&A...576A..15R} for long-term space photometry.

The new photometric period is longer by $\approx$10\%\ compared to that from Paper~I and is based on a data set that is roughly four times longer. The 2f harmonic of 13.9\,d in Paper~I also indicated  a longer period of around 28\,d. The second, more noisy part of the data between March 2007--December 2014 yields a 31.8\,d period but with a very low significance. Putting together all the available photometric data results in an $\approx$28\,d period. Therefore, we accept $P_{\rm rot}=28.30$\,d from the first data set as the most feasible and accurate rotation period. Accordingly, for phase calculations we use the following equation:
\begin{equation}\label{eq1}
{\rm HJD} = 2,450,369.2 + 28.30\times E,
\end{equation}
where the reference time was chosen arbitrarily (cf. Paper~I).

\section{Fundamental parameters}\label{astrop}

We redetermined the effective temperature ($T_{\rm eff}$), surface gravity ($\log g$), metallicity ([Fe/H]), and projected rotational velocity ($v\sin i$) via the spectrum-synthesis code ParSES
\citep{2004AN....325..604A,2013POBeo..92..169J} implemented in the standard STELLA-SES data reduction process \citep{2008SPIE.7019E..0LW}. For the synthetic spectra, we determined a microturbulence $\xi_{\rm mic}$ of 1.25\,\kms\ by following the empirical relation as was used in the Gaia-ESO survey \citep{2014A&A...564A.133J}. The radial-tangential macroturbulence $\xi_{\rm mac}$ of 3\,\kms\ was taken from \citet{1997PASP..109..514F}, but see also Paper~I. The resulting parameters with their internal standard deviations are listed in Table~\ref{astropars}. When compared to the previous values in Paper~I, the gravity $\log g$ and $v\sin i$ are only slightly different; these values are still within the small error boxes, but the effective temperature of 4305\,K is lower by $\approx$200\,K. However, the new, lower, value is in a better agreement with the colour-index temperature calibration by \citet{2011ApJS..193....1W} when taking $V-I_C=1\fm33$ from the long-term photometric data. In Paper~I $V-I=1\fm21$ was taken from the Hipparcos/Tycho catalogue, which would have been in accordance with a temperature of $\approx$4400\,K.

\begin{figure}[]
\includegraphics[width=\columnwidth]{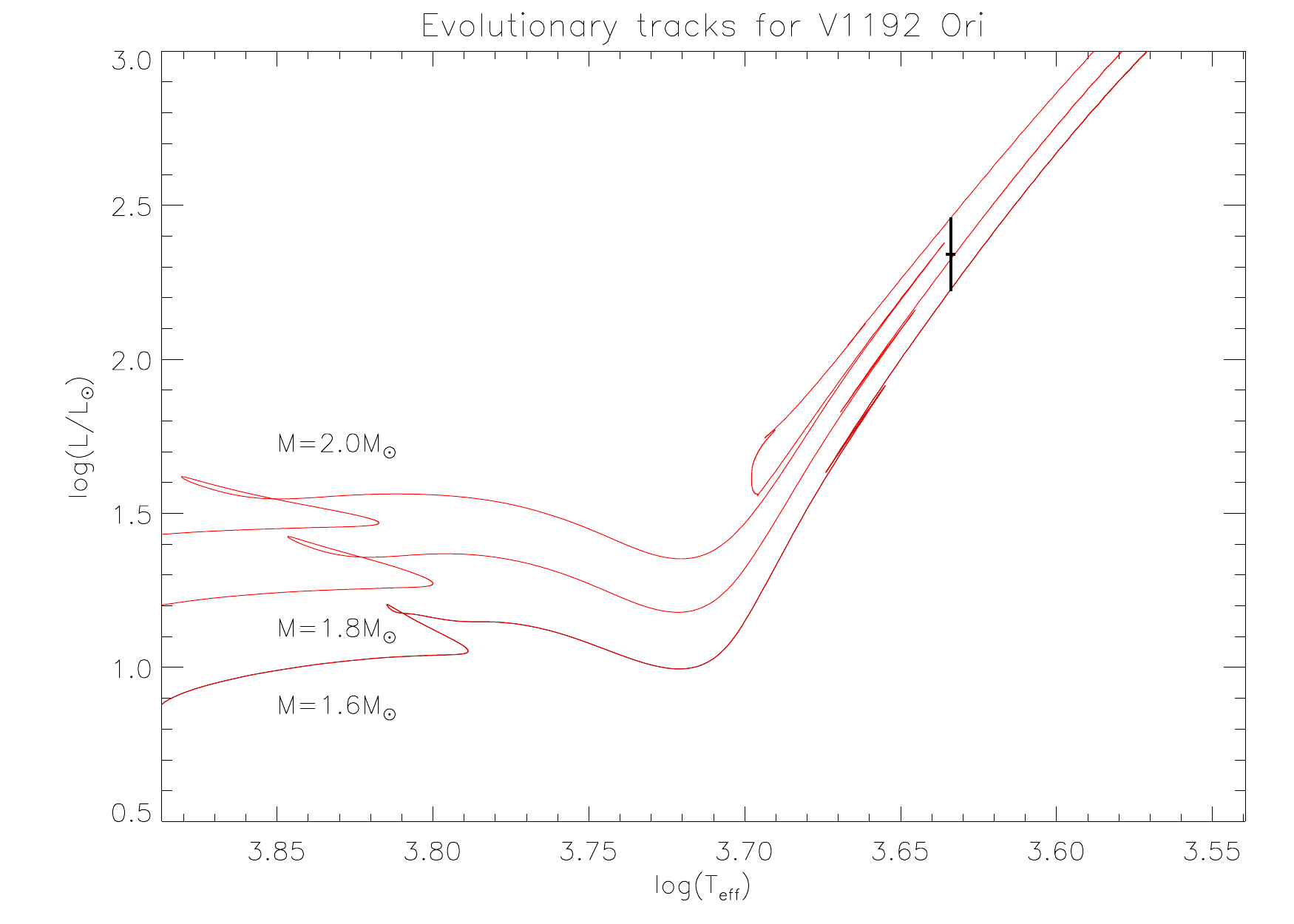}
\caption{Position of \vori (dot) in the H-R diagram. Shown are stellar evolutionary tracks for 1.6, 1.8, and 2.0\,M$_\odot$ from \emph{PARSEC}, assuming [Fe/H]=$-$0.08. The location of \vori at the RGB indicates a mass of 1.85\,$M_{\odot}$ and an age of 1.6\,Gyrs.}
\label{hrd}
\end{figure}

The photometric period of 28.3\,d together with the projected rotational velocity of 32.0$\pm$1.5\kms, and taking 65\degr inclination from Paper~I, yields a stellar radius $R=19.7^{+4.4}_{-2.2}$~R$_{\odot}$ which, together with $T_{\rm eff}=4305$\,K, fits fairly well for a K2.5\,III classification \citep{1996AJ....111.1705D,1999AJ....117..521V}. 
In  Paper~I a parameter study was carried out to select the true inclination by achieving the most homogeneous temperature inversion.  For our new spectroscopic data we carried out a similar test with our new inversion
code \emph{iMap}  (see Sect.~\ref{imap}) and obtained the most acceptable inversions for a range of inclination angles between 50\degr-70\degr. However, \emph{iMap} works in a different way and is only of limited use to perform such a parameter search. Thus, we decided to keep the former inclination angle from Paper~I assuming a slightly larger error bar of $\pm$15\degr.
From the radius and effective temperature it follows that the luminosity $L=120^{+61}_{-26}L_{\odot}$. When assuming $M_{{\rm bol,}\odot}=4\fm74$ this yields $M_{\rm bol}=-0\fm45^{+0.26}_{-0.46}$.

\begin{table}[]
\caption{Astrophysical properties of \vorid} \label{astropars}
 \centering
 \begin{tabular}{lll}
  \hline\noalign{\smallskip}
  Parameter               &  Value \\
  \hline\hline
  \noalign{\smallskip}
  Spectral type            & K2.5\,III  \\
  Distance$_{\rm Gaia}$ [pc]    &  $327^{+33}_{-28}$\\
  $V_{\rm br}$    [mag]          & $7\fm42\pm0\fm10$ \\
  $(V-I)_{C,{\rm br}}$  [mag]         & $1\fm33\pm0\fm03$ \\
  $M_{\rm bol}$     [mag]        & $-1\fm10^{+0.29}_{-0.32}$ \\
  Luminosity [${L_{\odot}}$]         & $218^{+73}_{-52}$ \\
  $\log g$ [cgs]        &          $ 2.39\pm0.04$ \\
  $T_{\rm eff}$ [K]           &           $4305\pm15 $    \\
  $v\sin i$ [km\,s$^{-1}$]           &         $32.0\pm1.5$ \\
  Rotation period [d]   &           $28.30\pm0.02 $ \\
  Inclination  [\degr]            &       $65\pm15$ \\
  Radius      [$R_{\odot}$]           &      $19.7^{+4.3}_{-2.1}$   \\
  Mass          [$M_{\odot}$]           & $1.85\pm0.30$   \\
  Age  [Gyr]                    &   $1.64\pm0.3$\\
  Microturbulence  [km\,s$^{-1}$] & $1.25$ \\ 
  Macroturbulence  [km\,s$^{-1}$] & 3.0 \\
  Metallicity [Fe/H] &  $-0.08\pm 0.02$ \\
  NLTE Li abundance (log)   & 1.27$\pm$0.15  \\
\hline
 \end{tabular}
\end{table}

\begin{figure}[]
\includegraphics[width=1.0\columnwidth]{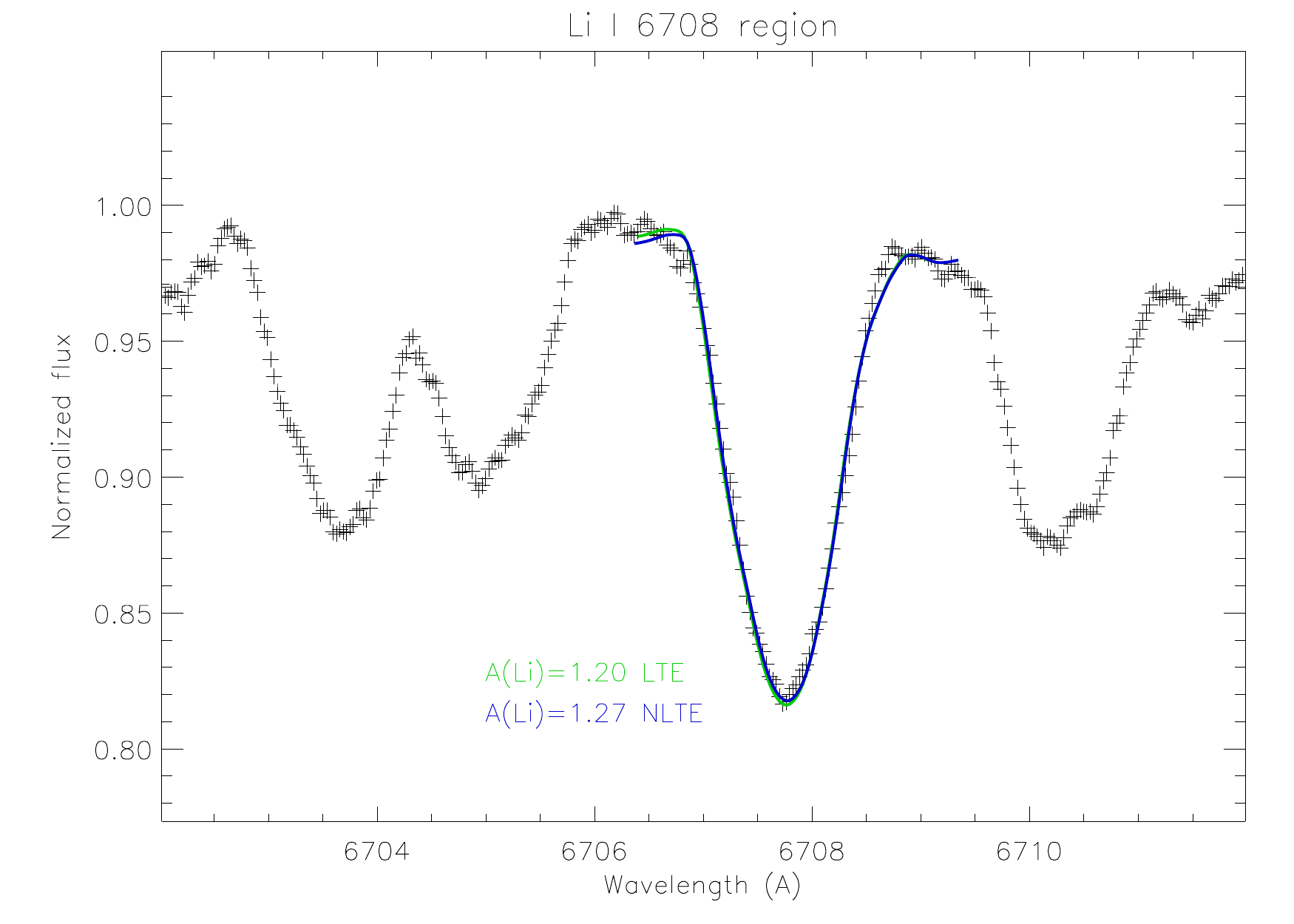}
\caption{Observed Li\,{\sc i} 6708\,\AA\ spectral region of \vori fitted by synthetic Li profiles using LTE (green) and NLTE (blue) approximation.}
\label{li_spectrum}
\end{figure}

The improved parallax of 3.06$\pm$0.28\,mas from \emph{Gaia} DR1 \citep{2016A&A...595A...1G,2016A&A...595A...2G,2016A&A...595A...4L} yields a distance of $327^{+33}_{-28}$\,pc, which is 25\%\ larger compared to the \emph{Hipparcos} distance of $238^{+83}_{-19}$ (cf. Paper~I).
Based on our long-term APT photometry we are able to give a new estimation of $7\fm42\pm0\fm10$ for the brightest $V$  magnitude  observed so far (see $V_{\rm br}$ in Table~\ref{astropars}). Taking these improved values and assuming an interstellar extinction of $A_V=0\fm191$ \citep{1998ApJ...500..525S} together with a bolometric correction of $BC=-0\fm76$ \citep{1996ApJ...469..355F} results in $M_{\rm bol}=-1\fm10^{+0.29}_{-0.32}$. 
This value and that calculated from $T_{\rm eff}$ and $R$ 
agree with each other within their errors. The value
$M_{\rm bol}$ derived from the \emph{Gaia} parallax converts to a luminosity of $218^{+73}_{-52}\,L_{\odot}$, which we eventually adopted to find the most plausible position of \vori in the H-R diagram in Fig.~\ref{hrd}. We adopted the \emph{PARSEC} stellar evolution grid by \citet{2012MNRAS.427..127B,2013EPJWC..4303001B}, interpolating for [Fe/H] of $-$0.08 (Z=0.0157). From the model grid we obtained a mass of 1.85$\pm$0.3\,$M_{\odot}$ with an age of 1.64$\pm$0.3\,Gyr, typical for the epoch just after the RGB bump. This new mass is consistent with the value of 1.9$\pm$0.3\,$M_{\odot}$ from Paper~I. Taking the mass and radius would yield $\log g=2.12^{+0.17}_{-0.27}$, i.e. a bit lower than the adopted value from ParSES. However, such a difference can originate from, for example a somewhat underestimated $V_{\rm br}$ \citep[cf.][]{2014A&A...572A..94O}, which would yield lower luminosity, therefore lower mass. Also, a 0.02 dex shift in metallicity yields a mass difference of $\approx$0.1.

\begin{table*}[]
 \centering
\caption{Temporal distribution of the Doppler images}
\label{disets}
\begin{tabular}{c c c c c c c }
\hline\noalign{\smallskip}
Observing & Doppler & Mid-HJD & Mid-date & Number & Data coverage & Data coverage \\
 run & image  &2\,450\,000+  & yyyy-mm-dd& of spectra & [days] & in $P_{\rm rot}$  \\
\hline
\hline
\noalign{\smallskip}
1st & S11  & 4154.266 & 2007-02-22  & 24 & 26.914 & 0.951  \\
\hline
\noalign{\smallskip}
2nd & S21  &   4353.335   & 2007-09-09 & 23 & 26.968 & 0.953 \\
  & S22  &   4380.154  & 2007-10-06  & 20 & 26.957 & 0.953 \\
 & S23 &  4413.098  &  2007-11-08 & 27 & 42.950 &  1.518 \\
 & S24  &   4470.907  & 2008-01-05 & 21 & 36.000 & 1.272\\
 & S25  &  4505.701  & 2008-02-09 & 19 & 45.969 & 1.624 \\
\hline
\noalign{\smallskip}
3rd & S31  &   4785.477 & 2008-11-14 & 22 & 26.951 & 0.952  \\
 & S32  &   4820.173  & 2008-12-19  & 17 & 30.844   & 1.090 \\
\hline
\noalign{\smallskip}
4th & S41  &  5146.236  &  2009-11-10  & 19 & 38.924  &  1.375 \\
\hline
\noalign{\smallskip}
5th & S51   &  5616.227  & 2011-02-23  & 12 & 37.919  &1.340  \\
\hline
\noalign{\smallskip}
6th & S61  &   7408.699   & 2016-01-21  & 25 & 26.025  &  0.920 \\
\hline
\end{tabular}
\end{table*}

\begin{figure*}[]
\vspace{2.5cm}{\hspace{0.5cm}\Large{S11}}
\vspace{-2.5cm}

\hspace{2.00cm}\includegraphics[angle=0,width=1.74\columnwidth]{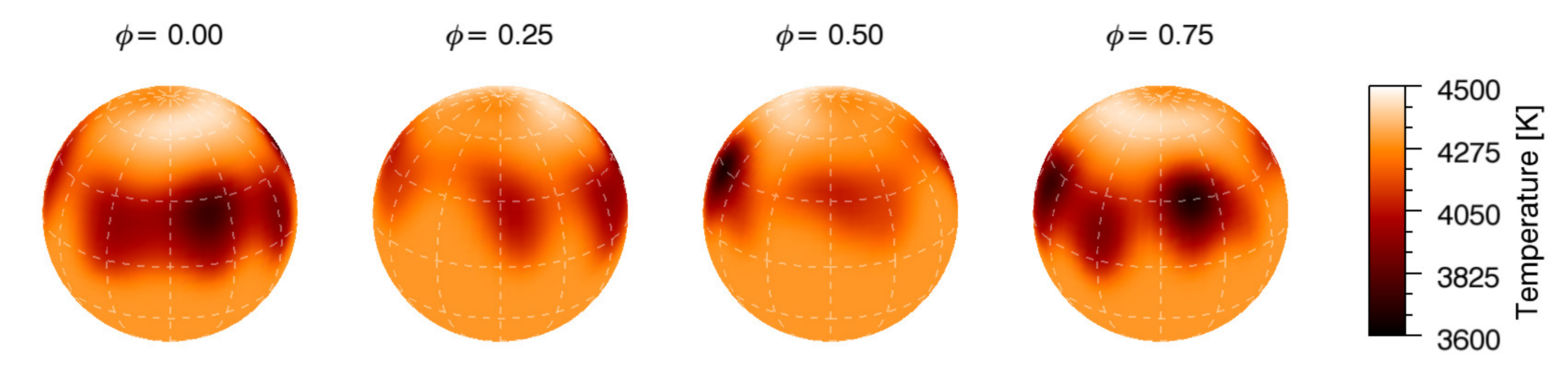}

\caption{Doppler image of \vori for the S11 data set. The corresponding mid-date is 2007-02-22.
The spherical surface map is shown in four rotational phases (identified on top) along with the temperature scale. }
\label{dis11}
\end{figure*}

\section{Surface Li abundance}\label{lithium}

In Fig.~\ref{li_spectrum} we plot an average Li\,{\sc i}-6708\,\AA\ spectrum from summing up seven good-quality (S/N ratio of $\approx$200) spectra distributed evenly along all rotation phases in 2008. This spectrum yields S/N>500:1. For the abundance determination, we first employed SME spectral synthesis \citep{2017A&A...597A..16P} with MARCS model atmospheres \citep{2008A&A...486..951G} and with LTE approximation. Atomic data were gathered from the Vienna Atomic Line Database (VALD) \citep{1999A&AS..138..119K}. The best fit resulted in $A(\mathrm{Li})=1.20$, i.e. somewhat lower than the value of 1.4$\pm$0.2 from  \citet{1993ApJ...403..708F}.  A 100\,K change in $T_{\rm eff}$ yields an $\approx$0.15\,dex uncertainty, while a 2\%\ change in the continuum level yields 0.10\,dex.

The NLTE synthesis was also carried out with SME using MARCS model atmospheres but with pre-computed departure coefficients. The fit resulted in $A(\mathrm{Li})_{\mathrm{NLTE}}=1.27$, in good agreement with the recent result by \citet{2015AJ....150..123R}. We plot the LTE and the NLTE fits together in Fig.~\ref{li_spectrum}. Both the LTE and the NLTE approach yields abundances lower than the anticipated limiting value of 1.5 for Li-rich giants. By applying the independent LTE-NLTE abundance correction of \citet{2016A&A...586A.156K} to our LTE abundance yields $A(\mathrm{Li})_{\mathrm{ NLTE}}=1.46$, i.e. also  below the nominal 1.5. Thus, we conclude that \vori is actually not a bona fide Li-rich giant but a Li-normal star with high surface amounts of Li.

\section{Doppler images for 2007--2016}\label{di}

\begin{figure*}[t]
\vspace{2.5cm}{\hspace{0.5cm}\Large{S21}}
\vspace{-2.5cm}

\hspace{2.00cm}\includegraphics[angle=0,width=1.74\columnwidth]{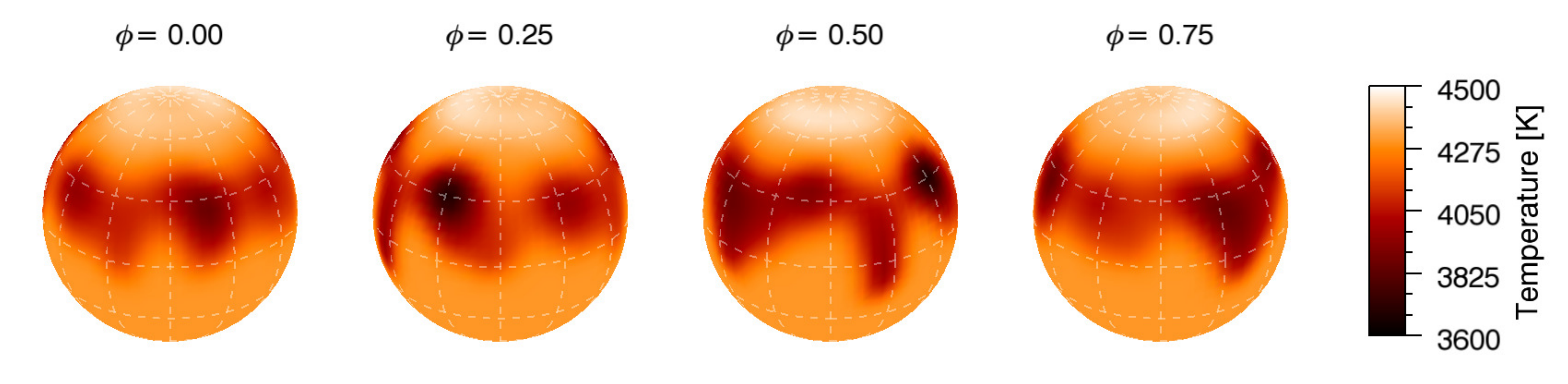}

\vspace{2.5cm}{\hspace{0.5cm}\Large{S22}}
\vspace{-2.5cm}

\hspace{2.00cm}\includegraphics[angle=0,width=1.74\columnwidth]{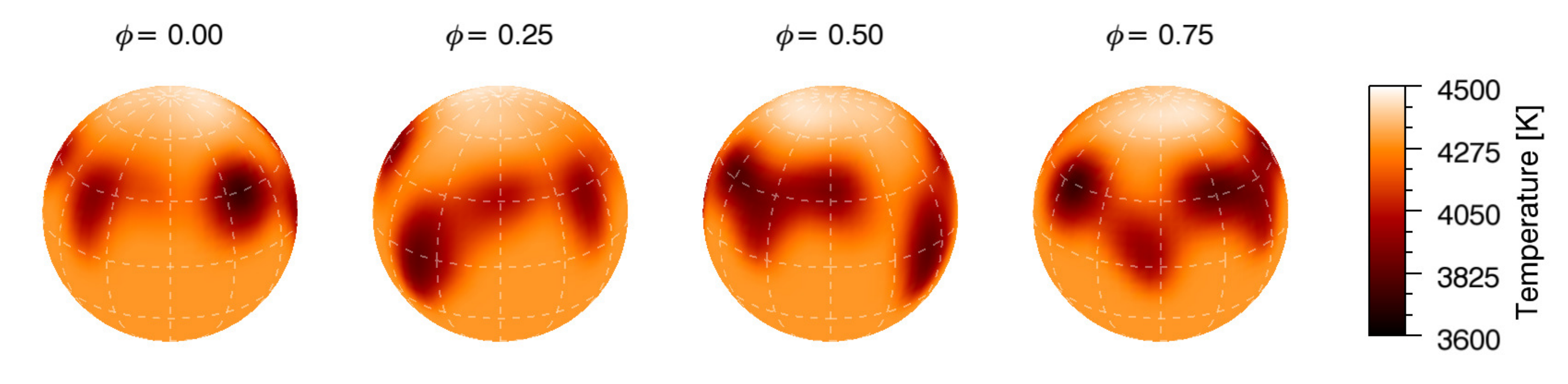}

\vspace{2.5cm}{\hspace{0.5cm}\Large{S23}}
\vspace{-2.5cm}

\hspace{2.00cm}\includegraphics[angle=0,width=1.74\columnwidth]{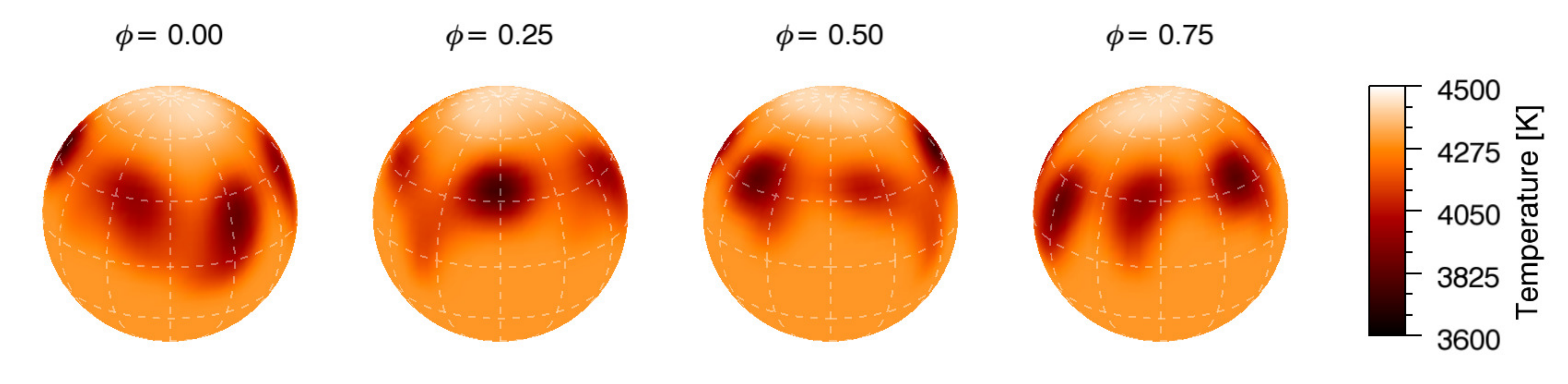}

\vspace{2.5cm}{\hspace{0.5cm}\Large{S24}}
\vspace{-2.5cm}

\hspace{2.00cm}\includegraphics[angle=0,width=1.74\columnwidth]{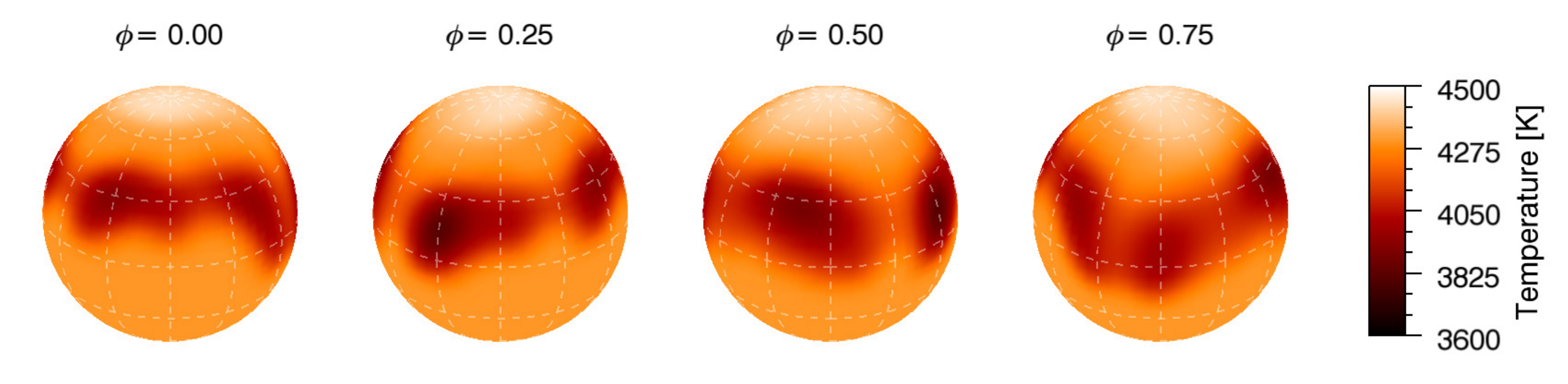}

\vspace{2.5cm}{\hspace{0.5cm}\Large{S25}}
\vspace{-2.5cm}

\hspace{2.00cm}\includegraphics[angle=0,width=1.74\columnwidth]{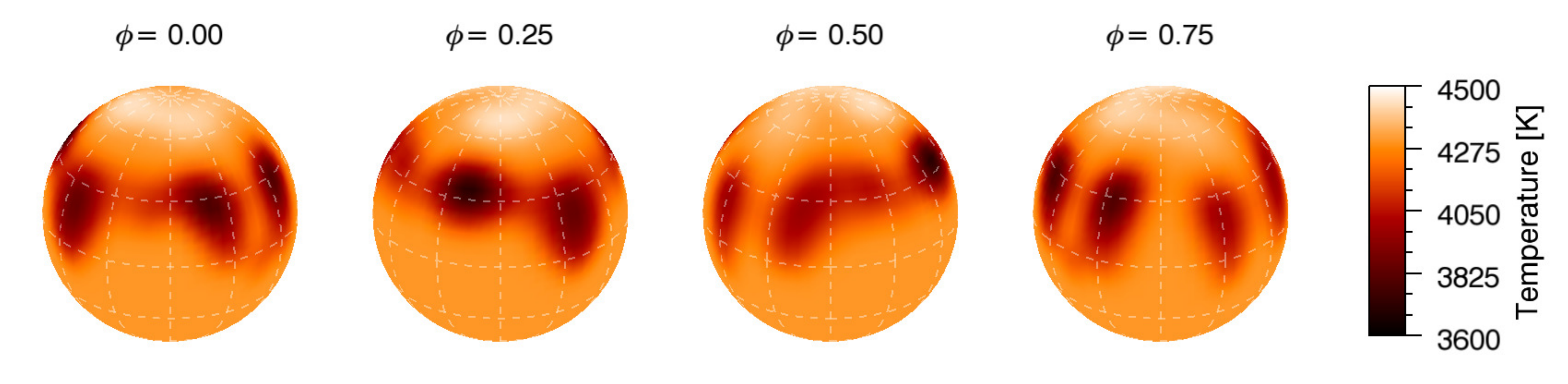}

\caption{Doppler images of \vori for the five data sets S21, S22, S23, S24, and S25. The corresponding mid-dates are
2007-09-09, 2007-10-06,  2007-11-08, 2008-01-05, and 2008-02-09 , respectively.
Otherwise as in Fig.~\ref{dis11}. }
\label{dis2x}
\end{figure*}

\subsection{Configuring data subsets for Doppler imaging}

Our spectroscopic data were taken during six observing runs between 2007 and 2016, each providing fairly good sampling for the relatively long rotational phase of 28.3~days. This data set allowed for  altogether 11 Doppler reconstructions. Table~\ref{disets} summarizes the temporal distribution of the Doppler reconstructions over the six runs (see also Table~\ref{Tab1} in the Appendix). The second and the third runs were long and continuous enough for obtaining Doppler images for several consecutive stellar rotations, suitable for studying surface differential rotation by tracking short-term spot displacements (see Sect.~\ref{ccf}).

\subsection{Image reconstruction code iMap}\label{imap}

Our Doppler reconstruction code \emph{iMap} performs multi-line inversion simultaneously for a large number of
photospheric line profiles \citep{2012A&A...548A..95C}. For the inversion we selected 40 suitable absorption lines from the 5000--6750 \AA\ wavelength range by their line depth,
blends, continuum level, and their temperature sensitivity \citep{2015A&A...578A.101K}. Each contributing line is modelled individually and locally and then disk-integrated; finally, all disk-integrated line profiles are averaged to form the mean line profile, which can be compared with each observed mean profile for each observed phase \citep[for more details see Sect. 3 in][]{2012A&A...548A..95C}.

The \emph{iMap} code calculates the line profiles by solving the radiative transfer through an artificial
neural network \citep{2008A&A...488..781C}. Atomic parameters are taken from the VALD database
\citep{1999A&AS..138..119K}. Model atmospheres are taken from \citet{2004astro.ph..5087C} and
are interpolated for each desired temperature, gravity, and metallicity. Owing to the high workload for computation and modelling, we used LTE radiative transfer instead of spherical model atmospheres. Nevertheless, limitations from neglecting spherical model atmospheres and continuum scattering are compensated by using dense phase coverages (cf. Table~\ref{disets}) and also by using our multi-line approach.
For the surface reconstruction \emph{iMap} uses an iterative regularization
based on a Landweber algorithm \citep{2012A&A...548A..95C}, and therefore no additional
constraints are imposed in the image domain. For the inversions we used the same stopping criteria as given by \citet{2012A&A...548A..95C}. According to our tests (see Appendix A in the aforementioned reference) the iterative regularization (i.e. step size control \& stopping rule) is proved to be enough to converge always to the same image solution. The surface element resolution is set to $5^{\circ} \times 5^{\circ}$.

\begin{figure*}[thb]
\vspace{2.5cm}{\hspace{0.5cm}\Large{S31}}
\vspace{-2.5cm}

\hspace{2.00cm}\includegraphics[angle=0,width=1.74\columnwidth]{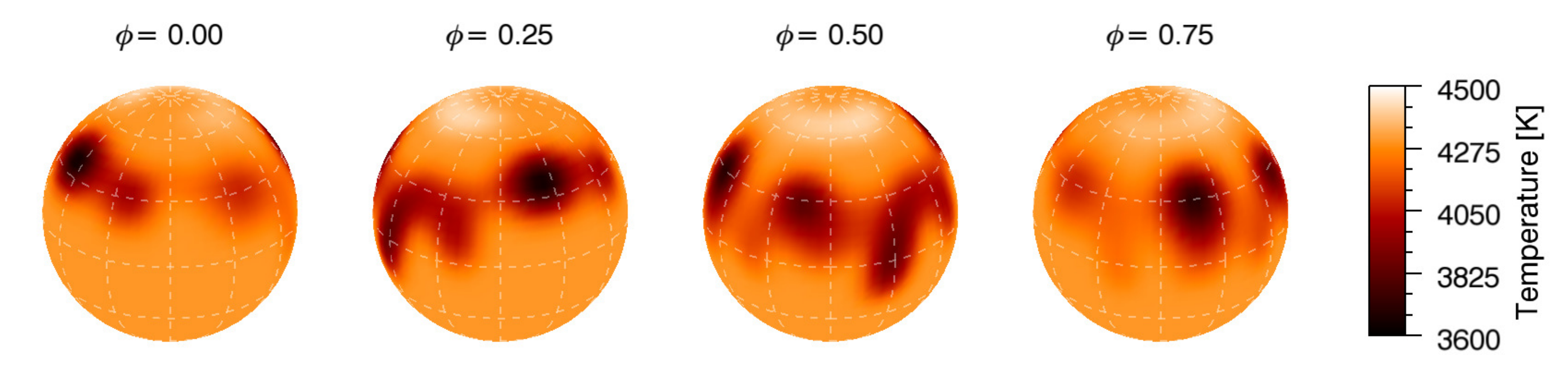}

\vspace{2.5cm}{\hspace{0.5cm}\Large{S32}}
\vspace{-2.5cm}

\hspace{2.00cm}\includegraphics[angle=0,width=1.74\columnwidth]{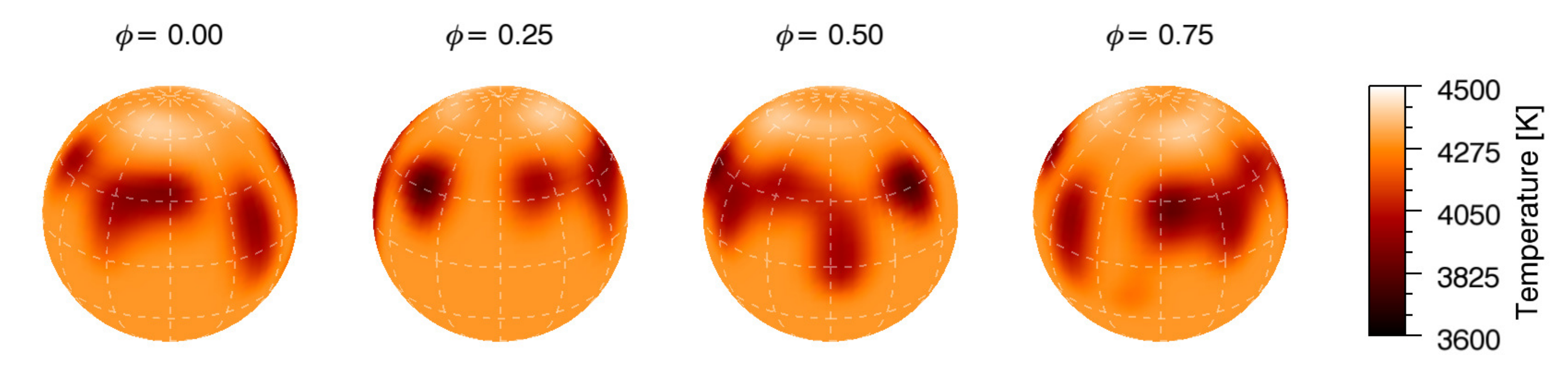}

\caption{Doppler images of \vori for the two data sets S31 and S32. The corresponding mid-dates are
2008-11-14 and 2008-12-19, respectively.
Otherwise as in Fig.~\ref{dis11}. }
\label{dis3x}
\end{figure*}

\begin{figure*}[thb]
\vspace{2.5cm}{\hspace{0.5cm}\Large{S41}}
\vspace{-2.5cm}

\hspace{2.00cm}\includegraphics[angle=0,width=1.74\columnwidth]{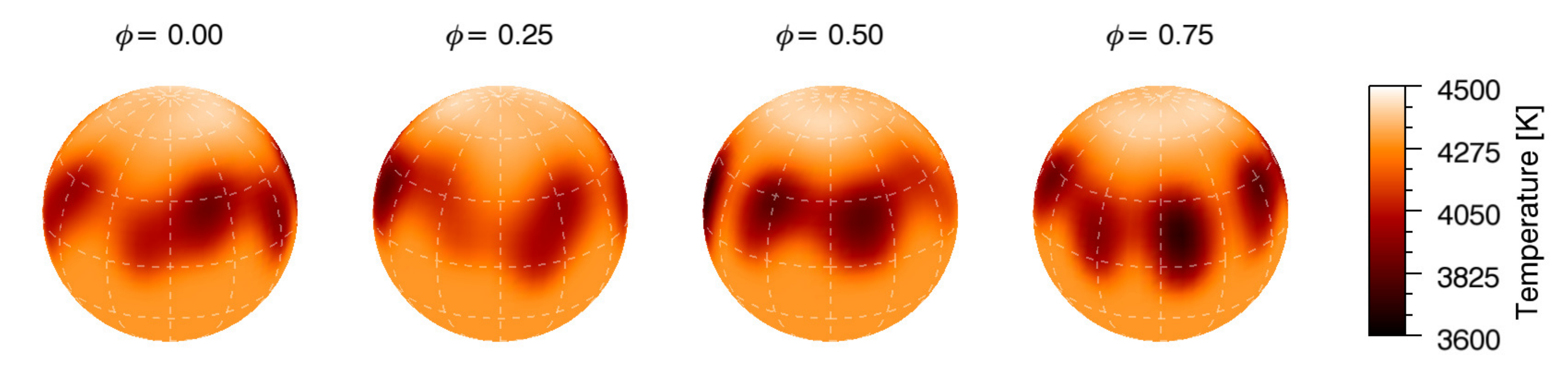}

\caption{Doppler image of \vori for the S41 data set. The corresponding mid-date is 2009-11-10.
Otherwise as in Fig.~\ref{dis11}.}
\label{dis41}
\end{figure*}

\begin{figure*}[thb]
\vspace{2.5cm}{\hspace{0.5cm}\Large{S51}}
\vspace{-2.5cm}

\hspace{2.00cm}\includegraphics[angle=0,width=1.74\columnwidth]{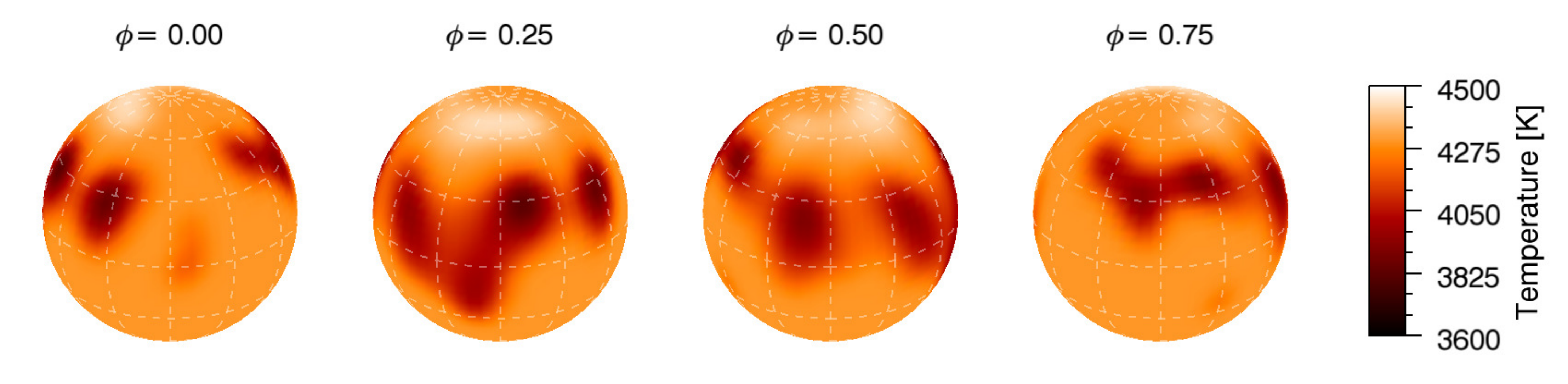}

\caption{Doppler image of \vori for the S51 data set. The corresponding mid-date is 2011-02-23.
Otherwise as in Fig.~\ref{dis11}. }
\label{dis51}
\end{figure*}

\begin{figure*}[thb]
\vspace{2.5cm}{\hspace{0.5cm}\Large{S61}}
\vspace{-2.5cm}

\hspace{2.00cm}\includegraphics[angle=0,width=1.74\columnwidth]{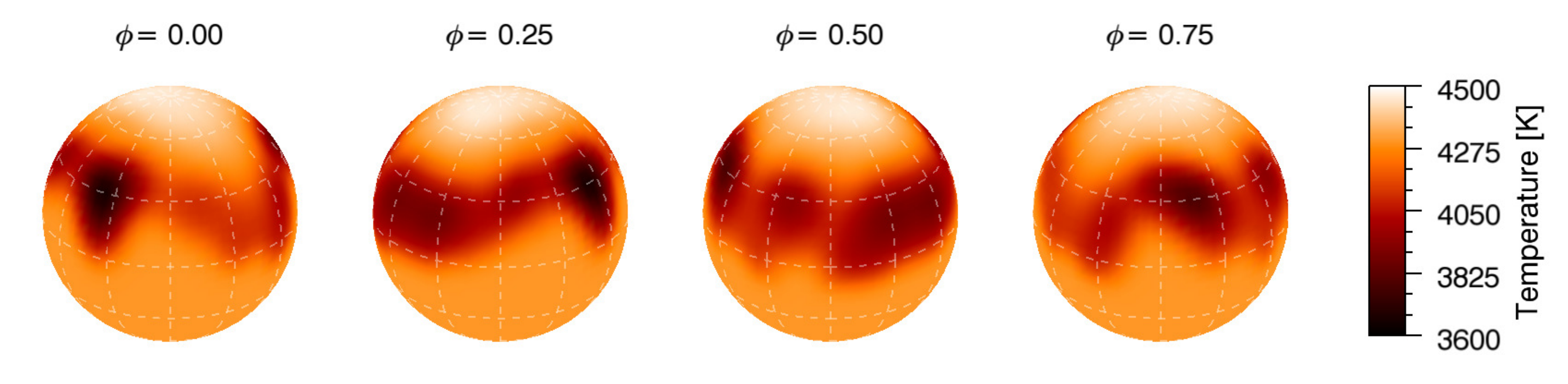}

\caption{Doppler image of \vori for the S61 data set. The corresponding mid-date is 2016-01-21.
Otherwise as in Fig.~\ref{dis11}.}
\label{dis61}
\end{figure*}

\subsection{Doppler image reconstructions}

Doppler reconstructions for \vori reveal a general characteristic; there are cool spots of different sizes and temperature contrasts all around the equatorial regions within a belt extending not higher than $\approx$50\degr\ and a warm azimuthal belt at higher latitudes or partly covering the pole. The  individual cool spots change in size considerably from one map to the next. The temperature of the coolest spots is $\approx$700\,K below the effective temperature of the unspotted photosphere. On the other hand, no cool spots at all appear on or near the visible pole during the time of observations, but the high latitude or even polar patches of $\approx$150\,K warmer regions are seen repeatedly. The overall surface structure with cool spots at lower latitudes and warmer but weakly contrasted features at higher latitudes bears resemblance with the first Doppler images from late 1996 in Paper~I, even though those maps revealed lower temperature contrasts. We evaluated the mean error for our temperature maps with a Monte Carlo analysis as described in \citet{2012A&A...548A..95C} and found a maximum error of 110\,K.

\emph{First run.} The very first season in early 2007 is covered by a single data set that allowed only one surface  reconstruction, shown in Fig.~\ref{dis11}. It reveals a chain of relatively large, partly adjoined spots all around the star. The individual spots have different contrasts but are always cooler than the effective temperature by $\approx$200--600\,K. The azimuthal-shaped warm feature extends half around the pole to a longitude that coincides with the most prominent cool spot at lower latitudes, implying that there may be a connection or that the warm feature may only be an artifact or at least raises doubts about its reality.

\emph{Second run.} The second season from late 2007 to early 2008 is our best-sampled season and provides five maps with a sampling of one map per stellar rotation. Again, as in early 2007, significant changes of the spot arrangement are seen from one map to the next. This is demonstrated in the time series of maps in  Fig.~\ref{dis2x}. Nevertheless, the corresponding dominant spots can easily be tracked on the consecutive maps and, besides some degree of sporadic displacements, the longitudinal tracks already indicate significant differential surface rotation (cf. Sect.~\ref{ccf}). This is particularly intriguing because the range of latitudes with spots as surface tracers is comparably narrow. There are also several other noteworthy morphological details, for example a persistent longitudinal gap between spots at around phase 0.4, or the one latitudinally displaced spot at ``southern'' latitude. At this point we point out that the Doppler-imaging technique would likely fail to resolve two close-together symmetric latitudinal belts of individual spots, for example as seen on our Sun. Simulations suggested that it likely reconstructs only a single belt placed at the sub-observers latitude \citep[e.g.][]{2000A&AS..147..151R}.

\emph{Third run.} For the third season in late 2008 (Fig.~\ref{dis3x}), we have another two consecutive maps. These maps again reveal spot rearrangements from one rotation to the next, which are partly morphological in nature and partly longitudinal migrations owing to differential surface rotation. The morphological changes were so rapid that there is almost no resemblance between the two maps; this is particularly the case, for example at phase 0.25 in Fig.~\ref{dis3x}, even though they are from two consecutive rotations. Such rapid variations are also seen on the Sun for particularly active spot groups, while solar plages may not even live as long as one solar rotation. Because the rotation periods of \vori and the Sun are not so different, 28.3~d versus 25~d, we may expect that some of the features we are mapping had evolved during the time of observation. If so, only a time average spot would be reconstructed.

\emph{Fourth, fifth, and sixth runs.} Finally, for the rest of the data, namely from late 2009 (Fig.~\ref{dis41}), early 2011 (Fig.~\ref{dis51}), and early 2016 (Fig.~\ref{dis61}) only one Doppler image per season was possible.  Yet, each one shows basically the same morphology, i.e. cool spots at low to mid-latitudes  distributed quasi-evenly along all rotational phases and weakly contrasted warm features at very high latitudes, but no cool spot at the pole itself.

\section{Surface differential rotation from time-series Doppler images}\label{ccf}

\begin{figure}[]
\includegraphics[width=1.0\columnwidth]{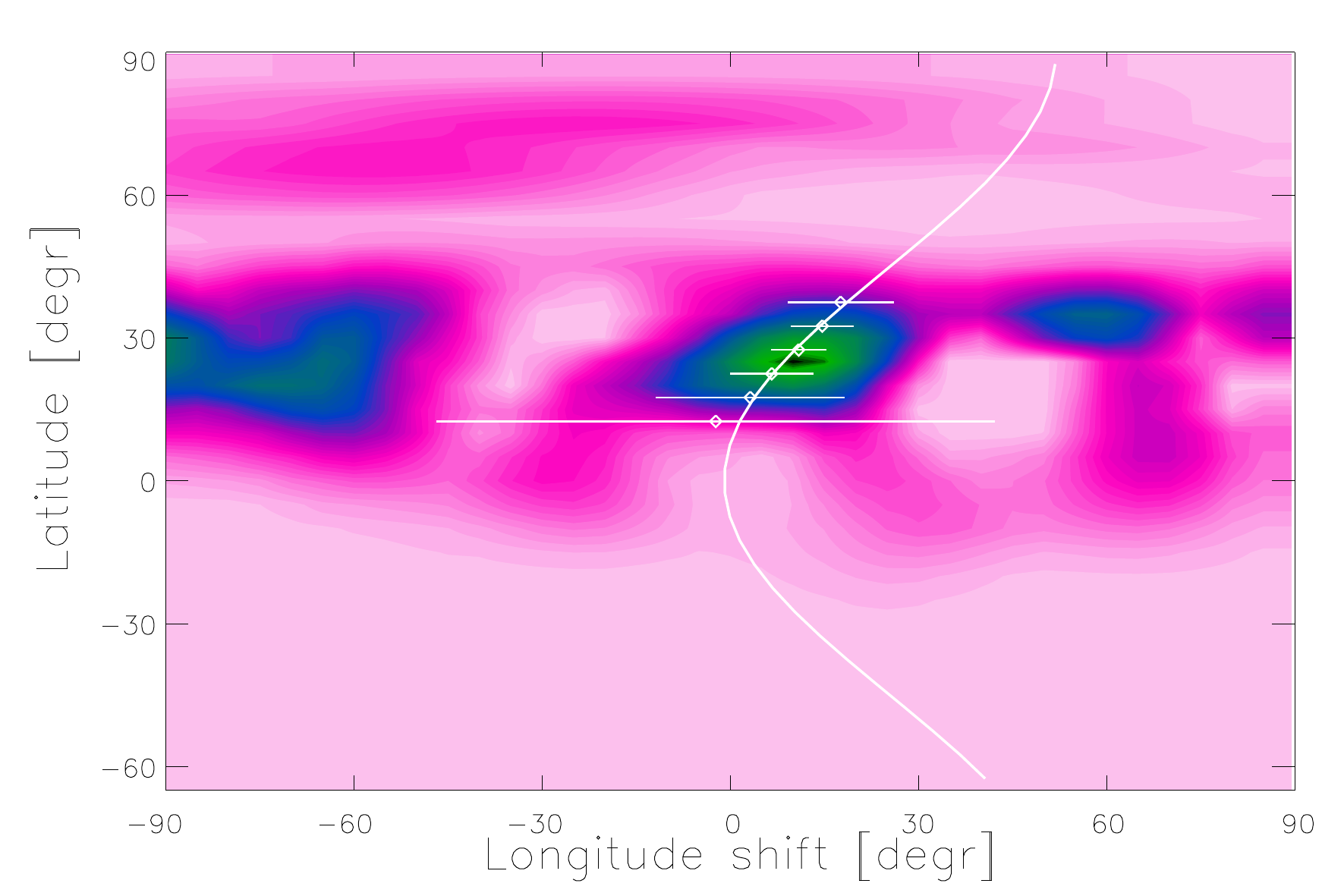}
\caption{Average cross-correlation map showing the evidence for surface differential rotation. Darker regions represent better correlation. The average longitudinal cross-correlation functions in 5\degr\ bins are fitted by Gaussian curves. Gaussian peaks are indicated by dots and the corresponding Gaussian widths by horizontal lines. The continuous line is the best fit, suggesting antisolar differential rotation with $P_{\rm eq}=28.35$\,d equatorial period and $\alpha=-0.11$ surface shear.}
\label{ccf_iMap}
\end{figure}

The time-series Doppler images in the second and third observing season allowed us to study the surface differential rotation by means of a cross-correlation analysis of the consecutive maps. Our cross-correlation technique \texttt{ACCORD} \citep[][and references therein]{2015A&A...573A..98K} combines the available information from spot displacements in order to reconstruct the signature of the differential rotation. In the second season from late 2007 to early 2008, we have five consecutive Doppler images (dubbed S21, S22, S23, S24, and S25), while in late 2008, we have two (S31 and S32). Therefore, we are able to create altogether five image pairs (S21-S22, S22-S23, S23-S24, S24-S25, and S31-S32) to be cross-correlated. These correlation maps are combined and the average correlation pattern is fitted with a 
quadratic rotation law. The result is shown in Fig.~\ref{ccf_iMap} and indicates  antisolar differential rotation, i.e. on \vori the rotation rate increases from the equator towards the pole. Its rotation law is expressed in the form $\Omega(\beta)=\Omega_{\rm eq}(1-\alpha\sin^2\beta)$, where $\Omega(\beta)$ is the angular velocity at $\beta$ latitude, $\Omega_{\rm eq}$ is the angular velocity at the equator, while $\alpha$ is the dimensionless surface shear coefficient obtained from $(\Omega_{\rm eq}-\Omega_{\rm pole})/\Omega_{\rm eq}$, i.e. the angular velocity difference between the equator and the pole divided by the equatorial velocity. The best fit yields $\Omega_{\rm eq}=12.695\pm0.034$\degr/d and $\alpha=-0.11\pm0.02$. This can be converted to a lap time of $\approx$260 days, that is the time the polar regions need to lap the equator by one full rotation.

\section{Summary and discussions}\label{disc}

\begin{table*}[thhh]
 \centering
\caption{Surface shear parameters obtained by cross-correlation of subsequent Doppler images}
\label{Tab5}
\begin{tabular}{l l l l  l }
\hline\noalign{\smallskip}
Star & Type  & $P_{\rm rot}$ [d]  & $\alpha$ surface shear & Reference \\
\hline
\hline
\noalign{\smallskip}
AB\,Dor         & K0V, single   & 0.51          & $+$0.005$\pm$0.001    &{\citet{1997MNRAS.291....1D}} \\
LQ\,Hya         & K2V, single   & 1.60          & $+$0.006$\pm$0.001    &{\citet{2004A&A...417.1047K}}\\
EI\,Eri                 & G5IV, binary  &  1.94         & $+$0.036$\pm$0.010    &{\citet{2009AIPC.1094..676K}} \\
UZ\,Lib                 & K0III,  binary        & 4.76          & $-$0.027$\pm$0.009             &{\citet{2007AN....328.1078V}}\\
HU\,Vir                 & K0IV, binary  & 10.39         & $-0.029\pm$0.005                       &{\citet{2016A&A...592A.117H}}\\
IL\,Hya                 & K0IV, binary  & 12.91         &  $+$0.05$\pm$0.01             &{\citet{2014IAUS..302..379K}}\\
DP\,CVn         & K1III, single & 14.01         &  $-$0.035$\pm$0.016   &{\citet{2013A&A...551A...2K}}\\
$\zeta$\,And    & K1III, binary & 17.76         &  $+$0.055$\pm$0.002   &{\citet{2012A&A...539A..50K}}\\
DI\,Psc                         & K1III, single & 18.02         &  $-$0.083$\pm$0.021    &{\citet{2014A&A...571A..74K}}\\
$\sigma$\,Gem   & K1III, binary & 19.60         &  $-$0.04$\pm$0.01             &{\citet{2015A&A...573A..98K}}\\
V2075\,Cyg      & K1III, binary & 22.62         &  $+$0.015$\pm$0.003   &{\citet{2016A&A...593A.123O}}\\
KU\,Peg                 & K2III, single & 23.90         &  $+$0.040$\pm$0.006   &{\citet{2016A&A...596A..53K}}\\
\vori                   & K2.5III, single       &28.30          &  $-$0.11$\pm$0.02             &this paper\\
\hline
\end{tabular}
\end{table*}
An extended Doppler imaging study with STELLA during the years 2007 to 2016 yielded 11 new surface image reconstructions, typically one image per stellar rotation. The 11 new Doppler images closely resemble the first Doppler image of \vori in Paper~I from 1996-97 taken with a different telescope-spectrograph combination and with a different inversion code. Its surface spot distribution is well characterized in the sense that cool spots of different sizes and temperatures are always found within a relatively confined latitudinal belt extending up to $\approx$50\degr\ and centred at the sub-observers latitude. No cool spots appear on or near the visible polar region. On the other hand, the individual spots change dynamically, not only from one observing season to the next but from one rotation to the next and possibly even within a single stellar rotation. A warm feature appears consistently at high latitudes as a (partial) azimuthal ring around the pole but we are not certain of its reality. Even though there are now maps spanning 20 years, one cannot identify a clear cyclic behaviour or trend. Long-term variability is certainly present in the photometric light curve on a timescale of 5--10 years (Fig.~\ref{aptdatafft}). However, the 10-yr gap in the photometric data from  March 1997 to  March 2007 prevents us from suggesting any cycle length. We nevertheless refine the rotation period of \vori from these data to 28.30~days and derive a more reliable set of fundamental stellar parameters by comparing to updated evolutionary tracks. Accordingly, the spectral type of \vori is found to be K2.5III, i.e. 0.5 subclasses cooler than claimed earlier.

For the time-series Doppler images from the years 2007--2008, we applied our robust cross-correlation technique and found strong antisolar differential rotation with an $\alpha=-0.11$ surface shear coefficient. This result is in very good agreement with the earlier result of $-0.12$ from the  independent data in Paper~I. This is the case even though the $\alpha$ value in Paper~I was derived only from a single cross-correlation of two consecutive Doppler maps, and therefore was significantly less robust and had larger error bars. With the present result, we are now confident that \vori indeed shows strong antisolar differential rotation. Together with DI\,Psc \citep{2014A&A...571A..74K} this is the strongest antisolar shear coefficient measured to date.
Such strong shear supports the relation of $|\alpha|\propto P_{\rm rot}$, i.e. the lower the rotation the stronger the shear \citep{2014SSRv..186..457K}. Also, a Rossby number of 0.22 derived for \vori \citep{2015A&A...574A..90A} indicates that an $\alpha\Omega$-type dynamo is operating underneath, i.e. differential rotation is expected to play an important role. 
Table~\ref{Tab5} compares \vori to other measurements of differential surface rotation. The list is not complete and, for the sake of homogeneity, only results from Doppler-imaging studies applying the cross-correlation technique are listed.

The unusually fast rotation of a single, evolved star such as \vori can be explained in various ways,  including angular momentum transport from the deep interior \citep{1989ApJ...346..303S}. The mass of 1.85\,$M_{\odot}$ found for \vori implies a precursor A5 spectral type on the main sequence \citep{1997A&A...327..207R} or even earlier in the case of a significant mass loss. Such a star does not have a deep outer convective zone to maintain a powerful magnetic dynamo that would result in effective magnetic braking on the main sequence \citep[cf.][]{2016A&A...591A..45P}. Eventually this means more angular momentum conservation for the post-main sequence evolution, supporting the scenario of mixing up angular momentum.  The position in the H-R diagram past the RGB luminosity bump (see Fig.\ref{hrd}) indicates that \vori has completed Li production at the red-giant bump. According to \citet{2000A&A...359..563C} the Li production is followed by an extra mixing phase, interconnecting the CN-burning zone with the convective envelope. Although the Li enrichment is relatively short lived, the extra mixing might explain the surface Li enrichment together with the peculiar rotation pattern. This is because freshly synthesized Li comes up to the surface along with high angular momentum material, which can eventually be conveyed towards the poles, resulting in the observed antisolar surface differential rotation \citep[cf.][]{2004AN....325..496K}.

This scenario is also compatible with the Cameron-Fowler mechanism \citep{1971ApJ...164..111C} of Li production together with the so-called cool-bottom processes \citep{1999ApJ...510..217S}, which are thought to be responsible for bringing down material from the convective zone and exposing that material to higher temperatures in which partial nuclear fusion (H burning) occurs. Then, the Li-rich material is transported back to the convective zone by some circulation or diffusion, wherefrom convective mixing spreads it out towards the surface \citep[see also][and their references]{2007ApJ...671..802B}. The competing scenario of planet engulfment \citep[cf.][their Sect.~6 and the references therein]{2017arXiv170406460M} explains the existence of the $^6$Li isotope in stellar atmospheres and may explain the rapid rotation as well, however, it would not easily account for antisolar differential rotation.

\begin{acknowledgements}
Authors thank the anonymous referee for his/her valuable comments and suggestions.  
We thank Dr.~Johanna Jurcsik for her helpful notes on deriving the correct rotation period by Fourier transformation. This paper is based on data obtained with the STELLA robotic telescopes in Tenerife, an AIP facility jointly operated by AIP and IAC (https://stella.aip.de/) and by the Amadeus APT jointly operated by AIP and Fairborn Observatory in Arizona. For their continuous support, we are grateful to the ministry for research and culture of the State of Brandenburg (MWFK) and the German federal ministry for education and research (BMBF).  Authors from Konkoly Observatory are grateful to the Hungarian National Research, Development and Innovation Office grants OTKA K-109276 and OTKA K-113117, and acknowledge support from the Austrian-Hungarian Action Foundation (OMAA). KV is supported by the Bolyai J\'anos Research Scholarship of the Hungarian Academy of Sciences.
The authors acknowledge the support of the German \emph{Deut\-sche For\-schungs\-ge\-mein\-schaft, DFG\/} through projects KO2320/1 and STR645/1.
This work has made use of data from the European Space Agency (ESA)
mission {\it Gaia} (\url{https://www.cosmos.esa.int/gaia}), processed by
the {\it Gaia} Data Processing and Analysis Consortium (DPAC,
\url{https://www.cosmos.esa.int/web/gaia/dpac/consortium}). Funding
for the DPAC has been provided by national institutions, in particular
the institutions participating in the {\it Gaia} Multilateral Agreement.
\end{acknowledgements}


\bibliography{kovarietal_v1192ori}
\bibliographystyle{aa}

%
%
\begin{appendix}
\section{Observing log of the spectroscopic data and line profile fits for each individual surface reconstruction}

\twocolumn
\onecolumn
\begin{center}
\longtab{
\begin{TableNotes}\footnotesize
\item[a]{2\,450\,000+}
\item[b]{Phases computed using Eq.~1.}
 \end{TableNotes}
\begin{longtable}{c c c c c|c c c c c}
\caption{\label{Tab1} Observing log of STELLA-SES spectra of \vori from 2007--2016 and grouping into subsets for Doppler reconstructions}\\
\hline\hline\noalign{\smallskip}
HJD\footnote  & Phase\footnote{b}\setcounter{footnote}{0} &  Date  & S/N & Subset  & HJD\footnote{a}  & Phase\footnote{b}\setcounter{footnote}{0} &  Date  & S/N & Subset  \\
\hline\noalign{\smallskip}
\endfirsthead
\caption{continued.}\\
\hline\hline\noalign{\smallskip}
HJD\footnote{a}   & Phase\footnote{b}\setcounter{footnote}{0} &  Date & S/N & Subset & HJD\footnote{a}   & Phase\footnote{b} &  Date & S/N & Subset\\
\hline\noalign{\smallskip}
\endhead
\hline
\insertTableNotes
\endfoot
\insertTableNotes
\endlastfoot
4140.464        &       0.260   &       08.02.2007      &       253     &       S11  &   4382.641   &       0.818   &       09.10.2007      &        147    &       S22\\
4141.454        &       0.295   &       09.02.2007      &       296     &       S11  &   4383.644   &       0.853   &       10.10.2007      &        203    &       S22\\
4142.454        &       0.331   &       10.02.2007      &       274     &       S11  &   4384.638   &       0.888   &       11.10.2007      &        132    &       S22\\
4143.455        &       0.366   &       11.02.2007      &       298     &       S11  &      4387.646        &       0.995   &       14.10.2007       &      186     &       S22\\
4144.447        &       0.401   &       12.02.2007      &       185     &       S11  &      4388.658        &       0.030   &       15.10.2007       &      182     &       S22\\
4146.456        &       0.472   &       14.02.2007      &       309     &       S11  &      4389.635        &       0.065   &       16.10.2007       &      183     &       S22\\
4147.414        &       0.506   &       15.02.2007      &       264     &       S11  &      4390.614        &       0.099   &       17.10.2007       &      74      &       S22\\
4148.397        &       0.541   &       16.02.2007      &       267     &       S11  &      4392.647        &       0.171   &       19.10.2007      &        182     &       S22\\
4150.417        &       0.612   &       18.02.2007      &       283     &       S11  &      4393.646        &       0.207   &       20.10.2007      &        140     &       S22\\
4152.458        &       0.684   &       20.02.2007      &       279     &       S11  &      4395.630        &       0.277   &       22.10.2007      &        123     &       S23\\
4153.458        &       0.719   &       21.02.2007      &       278     &       S11  &    4396.614  &       0.311   &       23.10.2007      &        137    &       S23\\
4154.476        &       0.755   &       22.02.2007      &       252     &       S11  &    4398.579  &       0.381   &       25.10.2007      &        126    &       S23\\
4155.480        &       0.791   &       23.02.2007      &       275     &       S11  &     4399.574 &       0.416   &       26.10.2007      &        125    &       S23\\
4156.459        &       0.825   &       24.02.2007      &       186     &       S11  &     4400.698 &       0.456   &       27.10.2007      &        122    &       S23\\
4157.460        &       0.861   &       25.02.2007      &       158     &       S11  &      4402.601        &       0.523   &       29.10.2007       &      164     &       S23\\
4158.460        &       0.896   &       26.02.2007      &       253     &       S11  &      4403.584        &       0.558   &       30.10.2007       &      110     &       S23\\
4159.459        &       0.931   &       27.02.2007      &       240     &       S11  &      4404.677        &       0.596   &       31.10.2007       &      162     &       S23\\
4160.461        &       0.967   &       28.02.2007      &       237     &       S11  &      4405.591        &       0.629   &       01.11.2007       &      177     &       S23\\
4162.384        &       0.035   &       02.03.2007      &       323     &       S11  &      4406.607        &       0.665   &       02.11.2007       &      151     &       S23\\
4163.361        &       0.069   &       03.03.2007      &       262     &       S11  &      4412.639        &       0.878   &       08.11.2007       &      57      &       S23\\
4164.389        &       0.106   &       04.03.2007      &       304     &       S11  &      4412.711        &       0.880   &       08.11.2007       &      79      &       S23\\
4165.375        &       0.140   &       05.03.2007      &       263     &       S11  &      4413.659        &       0.914   &       09.11.2007       &      156     &       S23\\
4166.378        &       0.176   &       06.03.2007      &       294     &       S11  &      4414.550        &       0.945   &       10.11.2007       &      92      &       S23\\
4167.378        &       0.211   &       07.03.2007      &       303     &       S11  &      4414.704        &       0.951   &       10.11.2007       &      121     &       S23\\
4338.707        &       0.265   &       26.08.2007      &       144     &       S21  &      4416.545        &       0.016   &       12.11.2007       &      52      &       S23\\
4339.684        &       0.300   &       27.08.2007      &       136     &       S21  &      4416.622        &       0.018   &       12.11.2007       &      174     &       S23\\
4340.684        &       0.335   &       28.08.2007      &       101     &       S21  &      4417.562        &       0.052   &       13.11.2007       &      126     &       S23\\
4343.685        &       0.441   &       31.08.2007      &       145     &       S21  &      4417.687        &       0.056   &       13.11.2007       &      121     &       S23\\
4345.673        &       0.511   &       02.09.2007      &       131     &       S21  &      4418.583        &       0.088   &       14.11.2007       &      117     &       S23\\
4346.672        &       0.547   &       03.09.2007      &       137     &       S21  &      4419.578        &       0.123   &       15.11.2007       &      141     &       S23\\
4347.674        &       0.582   &       04.09.2007      &       122     &       S21  &      4420.574        &       0.158   &       16.11.2007       &      146     &       S23\\
4349.697        &       0.654   &       06.09.2007      &       151     &       S21  &      4421.578        &       0.194   &       17.11.2007       &      144     &       S23\\
4350.699        &       0.689   &       07.09.2007      &       151     &       S21  &      4422.683        &       0.233   &       18.11.2007       &      148     &       S23\\
4351.692        &       0.724   &       08.09.2007      &       141     &       S21  &      4423.575        &       0.264   &       19.11.2007       &      177     &       S23\\
4352.693        &       0.759   &       09.09.2007      &       162     &       S21  &      4437.679        &       0.763   &       03.12.2007       &      220     &       S23\\
4353.693        &       0.795   &       10.09.2007      &       129     &       S21  &      4438.580        &       0.794   &       04.12.2007       &      327     &       S23\\
4354.688        &       0.830   &       11.09.2007      &       145     &       S21  &      4450.470        &       0.214   &       15.12.2007       &      395     &       S24\\
4355.692        &       0.865   &       12.09.2007      &       155     &       S21  &      4459.403        &       0.530   &       24.12.2007       &      271     &       S24\\
4357.683        &       0.936   &       14.09.2007      &       156     &       S21  &      4460.401        &       0.565   &       25.12.2007       &      330     &       S24\\
4358.695        &       0.972   &       15.09.2007      &       107     &       S21  &      4461.402        &       0.601   &       26.12.2007       &      352     &       S24\\
4359.672        &       0.006   &       16.09.2007      &       153     &       S21  &      4462.394        &       0.636   &       27.12.2007       &      342     &       S24\\
4360.673        &       0.041   &       17.09.2007      &       137     &       S21  &      4463.399        &       0.671   &       28.12.2007       &      327     &       S24\\
4361.669        &       0.077   &       18.09.2007      &       133     &       S21  &      4464.408        &       0.707   &       29.12.2007       &      356     &       S24\\
4362.669        &       0.112   &       19.09.2007      &       153     &       S21  &      4465.388        &       0.742   &       30.12.2007       &      351     &       S24\\
4363.672        &       0.147   &       20.09.2007      &       144     &       S21  &      4466.386        &       0.777   &       31.12.2007       &      344     &       S24\\
4364.667        &       0.183   &       21.09.2007      &       131     &       S21  &      4467.406        &       0.813   &       01.01.2008       &      306     &       S24\\
4365.675        &       0.218   &       22.09.2007      &       122     &       S21  &      4468.379        &       0.847   &       02.01.2008       &      328     &       S24\\
4366.689        &       0.254   &       23.09.2007      &       137     &       S22  &      4469.589        &       0.890   &       04.01.2008       &      263     &       S24\\
4367.650        &       0.288   &       24.09.2007      &       55      &       S22  &      4471.443        &       0.956   &       05.01.2008       &      319     &       S24\\
4369.656        &       0.359   &       26.09.2007      &       108     &       S22  &      4475.452        &       0.097   &       09.01.2008       &      357     &       S24\\
4370.684        &       0.395   &       27.09.2007      &       149     &       S22  &      4479.446        &       0.238   &       13.01.2008       &      315     &       S24\\
4371.702        &       0.431   &       28.09.2007      &       166     &       S22  &      4481.448        &       0.309   &       15.01.2008       &      334     &       S24\\
4372.664        &       0.465   &       29.09.2007      &       164     &       S22  &      4482.456        &       0.345   &       16.01.2008       &      325     &       S24\\
4373.664        &       0.501   &       30.09.2007      &       139     &       S22  &      4483.452        &       0.380   &       17.01.2008       &      344     &       S24\\
4377.662        &       0.642   &       04.10.2007      &       138     &       S22  &      4484.418        &       0.414   &       18.01.2008       &      203     &       S24\\
4378.653        &       0.677   &       05.10.2007      &       126     &       S22  &      4485.438        &       0.450   &       19.01.2008       &      182     &       S24\\
4379.649        &       0.712   &       06.10.2007      &       158     &       S22  &      4486.470        &       0.487   &       20.01.2008       &      274     &       S24\\
4380.641        &       0.747   &       07.10.2007      &       161     &       S22  &      4488.410        &       0.555   &       22.01.2008       &      89      &       S25\\
4488.481        &       0.558   &       22.01.2008      &       216      &       S25 & 5122.662  &       0.967   &       18.10.2009      &       299      &       S41\\
4490.497        &       0.629   &       24.01.2008      &       308 &   S25 & 5124.655      &       0.037   &       20.10.2009      &       149      &       S41\\
4491.452        &       0.663   &       25.01.2008      &       270 &   S25 & 5134.656      &       0.391   &       30.10.2009      &       292      &       S41\\
4492.389        &       0.696   &       26.01.2008      &       67      &       S25 & 5135.626      &       0.425   &       31.10.2009      &       336      &       S41\\
4494.422        &       0.768   &       28.01.2008      &       280     &       S25 & 5139.602      &       0.565   &       04.11.2009      &       318      &       S41\\
4495.407        &       0.802   &       29.01.2008      &       242     &       S25 & 5140.564      &       0.599   &       05.11.2009      &       126      &       S41\\
4496.454        &       0.839   &       30.01.2008      &       249     &       S25 & 5142.595      &       0.671   &       07.11.2009      &       270      &       S41\\
4499.444        &       0.945   &       02.02.2008      &       257     &       S25 & 5143.599      &       0.707   &       08.11.2009      &       250      &       S41\\
4501.456        &       0.016   &       04.02.2008      &       312     &       S25 &   5144.595    &       0.742   &       09.11.2009      &        273    &       S41\\
4502.457        &       0.052   &       05.02.2008      &       298     &       S25 &       5146.583        &       0.812   &       11.11.2009       &      323     &       S41\\
4508.460        &       0.264   &       11.02.2008      &       250     &       S25  &      5147.555        &       0.846   &       12.11.2009       &      146     &       S41\\
4512.461        &       0.405   &       15.02.2008      &       235     &       S25  &      5149.594        &       0.919   &       14.11.2009       &      225     &       S41\\
4514.462        &       0.476   &       17.02.2008      &       263     &       S25  &      5154.592        &       0.095   &       19.11.2009       &      251     &       S41\\
4515.461        &       0.511   &       18.02.2008      &       239     &       S25  &      5155.602        &       0.131   &       20.11.2009       &      279     &       S41\\
4516.461        &       0.546   &       19.02.2008      &       220     &       S25   &   5156.583  &       0.165   &       21.11.2009      &        159    &       S41\\
4532.378        &       0.109   &       06.03.2008      &       326     &       S25  &      5157.589        &       0.201   &       22.11.2009       &      268     &       S41\\
4533.379        &       0.144   &       07.03.2008      &       253     &       S25   &   5159.653  &       0.274   &       24.11.2009      &        315    &       S41\\
4534.379        &       0.179   &       08.03.2008      &       305     &       S25  &      5160.591        &       0.307   &       25.11.2009       &      307     &       S41\\
4774.642        &       0.669   &       04.11.2008      &       220     &       S31  &      5161.586        &       0.342   &       26.11.2009       &      256     &       S41\\
4775.595        &       0.703   &       05.11.2008      &       97      &       S31  &      5601.457        &       0.885   &       08.02.2011       &      72      &       S51\\
4775.639        &       0.705   &       05.11.2008      &       128     &       S31  &      5603.407        &       0.954   &       10.02.2011       &      96      &       S51\\
4775.726        &       0.708   &       05.11.2008      &       197     &       S31  &      5607.468        &       0.098   &       14.02.2011       &      107     &       S51\\
4776.644        &       0.740   &       06.11.2008      &       92      &       S31  &      5612.403        &       0.272   &       19.02.2011       &      114     &       S51\\
4776.733        &       0.743   &       06.11.2008      &       117     &       S31  &      5615.377        &       0.377   &       22.02.2011       &      115     &       S51\\
4777.595        &       0.774   &       07.11.2008      &       220     &       S31  &      5616.374        &       0.412   &       23.02.2011       &      113     &       S51\\
4778.595        &       0.809   &       08.11.2008      &       209     &       S31  &      5617.370        &       0.448   &       24.02.2011       &      139     &       S51\\
4779.590        &       0.844   &       09.11.2008      &       283     &       S31  &      5618.370        &       0.483   &       25.02.2011       &      110     &       S51\\
4780.586        &       0.879   &       10.11.2008      &       264     &       S31  &      5619.377        &       0.519   &       26.02.2011       &      91      &       S51\\
4781.670        &       0.918   &       11.11.2008      &       130     &       S31  &      5620.379        &       0.554   &       27.02.2011       &      59      &       S51\\
4783.586        &       0.985   &       13.11.2008      &       208     &       S31  &      5623.371        &       0.660   &       02.03.2011       &      122     &       S51\\
4784.581        &       0.021   &       14.11.2008      &       260     &       S31  &      5639.376        &       0.225   &       18.03.2011       &      94      &       S51\\
4785.581        &       0.056   &       15.11.2008      &       256     &       S31  &      7396.439        &       0.312   &       08.01.2016       &      182     &       S61\\
4789.583        &       0.197   &       19.11.2008      &       240     &       S31 &       7397.478        &       0.349   &       09.01.2016       &      387     &       S61\\
4793.594        &       0.339   &       23.11.2008      &       222     &       S31 &       7398.456        &       0.384   &       10.01.2016       &      364     &       S61\\
4794.588        &       0.374   &       24.11.2008      &       229     &       S31  &      7399.476        &       0.420   &       11.01.2016       &      359     &       S61\\
4795.584        &       0.409   &       25.11.2008      &       232     &       S31  &      7400.476        &       0.455   &       12.01.2016       &      372     &       S61\\
4798.671        &       0.518   &       28.11.2008      &       228     &       S31  &      7401.474        &       0.490   &       13.01.2016       &      196     &       S61\\
4799.533        &       0.549   &       28.11.2008      &       225     &       S31  &      7402.526        &       0.527   &       14.01.2016       &      324     &       S61\\
4800.593        &       0.586   &       30.11.2008      &       252     &       S31  &      7403.585        &       0.565   &       16.01.2016       &      200     &       S61\\
4801.593        &       0.622   &       01.12.2008      &       214     &       S31  &      7404.503        &       0.597   &       16.01.2016       &      350     &       S61\\
4805.597        &       0.763   &       05.12.2008      &       218     &       S32  &      7406.523        &       0.669   &       18.01.2016       &      365     &       S61\\
4808.589        &       0.869   &       08.12.2008      &       232     &       S32  &      7407.477        &       0.702   &       19.01.2016       &      406     &       S61\\
4809.663        &       0.907   &       09.12.2008      &       245     &       S32  &      7408.503        &       0.739   &       20.01.2016       &      395     &       S61\\
4810.591        &       0.940   &       10.12.2008      &       173     &       S32  &      7409.461        &       0.772   &       21.01.2016       &      361     &       S61\\
4811.596        &       0.975   &       11.12.2008      &       222     &       S32  &      7410.461        &       0.808   &       22.01.2016       &      326     &       S61\\
4812.595        &       0.010   &       12.12.2008      &       183     &       S32  &      7411.463        &       0.843   &       23.01.2016       &      370     &       S61\\
4814.603        &       0.081   &       14.12.2008      &       164     &       S32  &      7412.485        &       0.879   &       24.01.2016       &      377     &       S61\\
4817.606        &       0.187   &       17.12.2008      &       220     &       S32  &      7413.499        &       0.915   &       25.01.2016       &      364     &       S61\\
4818.494        &       0.219   &       17.12.2008      &       246     &       S32  &      7414.464        &       0.949   &       26.01.2016       &      405     &       S61\\
4819.470        &       0.253   &       18.12.2008      &       141     &       S32  &      7415.463        &       0.985   &       27.01.2016       &      390     &       S61\\
4822.479        &       0.360   &       21.12.2008      &       234     &       S32 &    7416.489   &       0.021   &       28.01.2016      &        187    &       S61\\
4827.430        &       0.535   &       26.12.2008      &       65      &       S32  &      7418.476        &       0.091   &       30.01.2016       &      361     &       S61\\
4828.460        &       0.571   &       27.12.2008      &       229     &       S32  &      7419.485        &       0.127   &       31.01.2016       &      242     &       S61\\
4829.451        &       0.606   &       28.12.2008      &       147     &       S32  &      7420.453        &       0.161   &       01.02.2016       &      69      &       S61\\
4834.442        &       0.782   &       02.01.2009      &       293     &       S32  &      7421.441        &       0.196   &       02.02.2016       &      195     &       S61\\
4835.440        &       0.818   &       03.01.2009      &       236     &       S32  &      7422.464        &       0.232   &       03.02.2016       &      277     &       S61\\
4836.441        &       0.853   &       04.01.2009      &       232     &       S32  &              &               &               &               &       \\
\hline
\end{longtable}
}
\end{center}

\twocolumn


\begin{figure*}[tb]
\centering
\vspace{0.35cm}
\hspace{-2.1cm}\large{S11}\hspace{5.47cm}\large{S21}\hspace{5.47cm}\large{S22}
\vspace{-0.35cm}

\includegraphics[width=0.565\columnwidth]{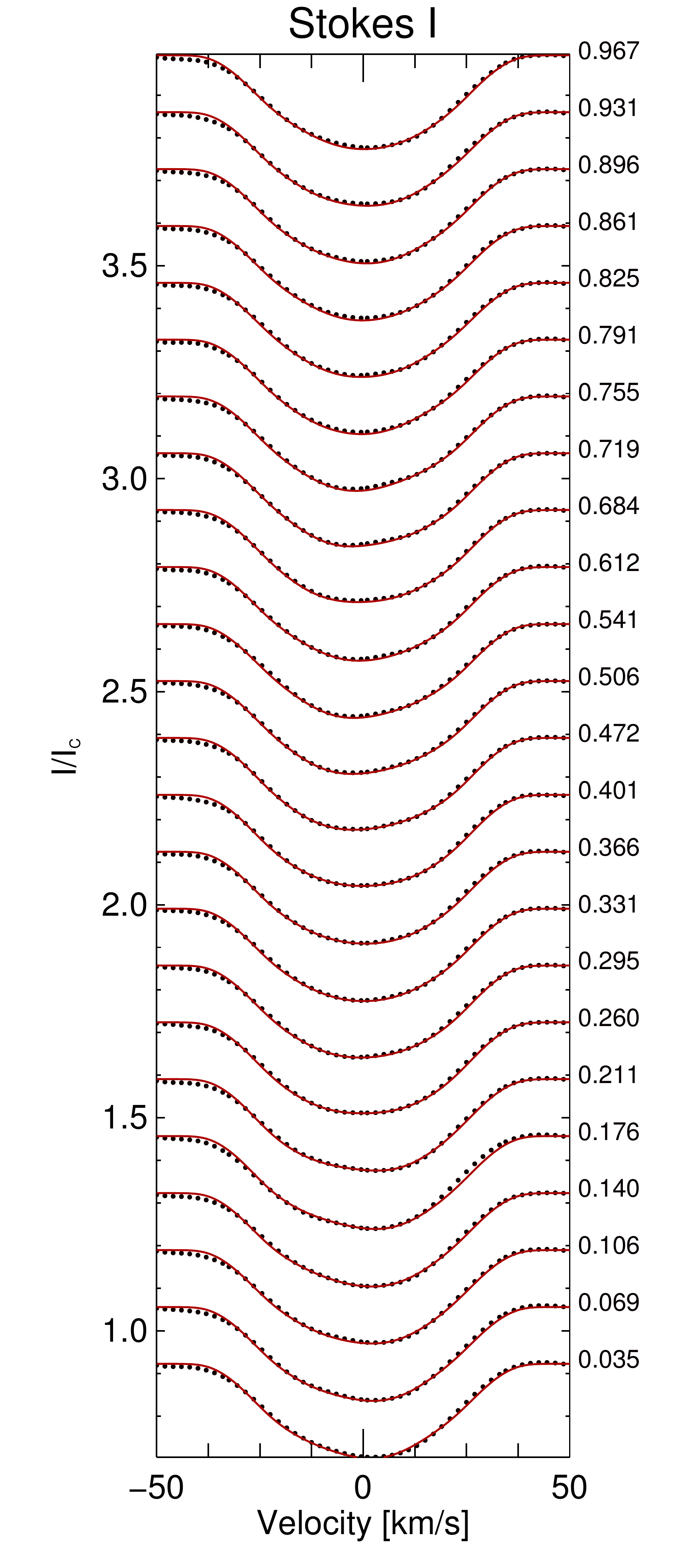}\hspace{9mm}
\includegraphics[width=0.565\columnwidth]{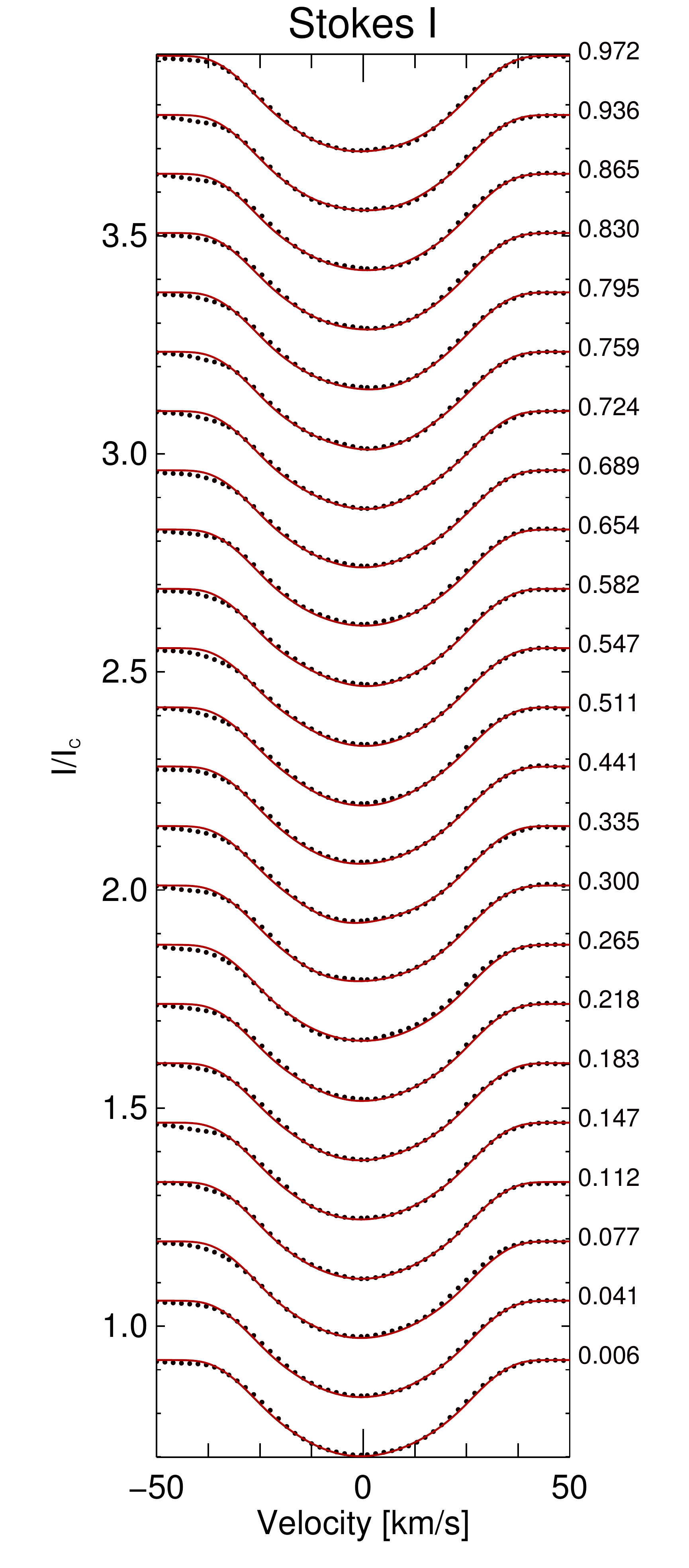}\hspace{9mm}
\includegraphics[width=0.565\columnwidth]{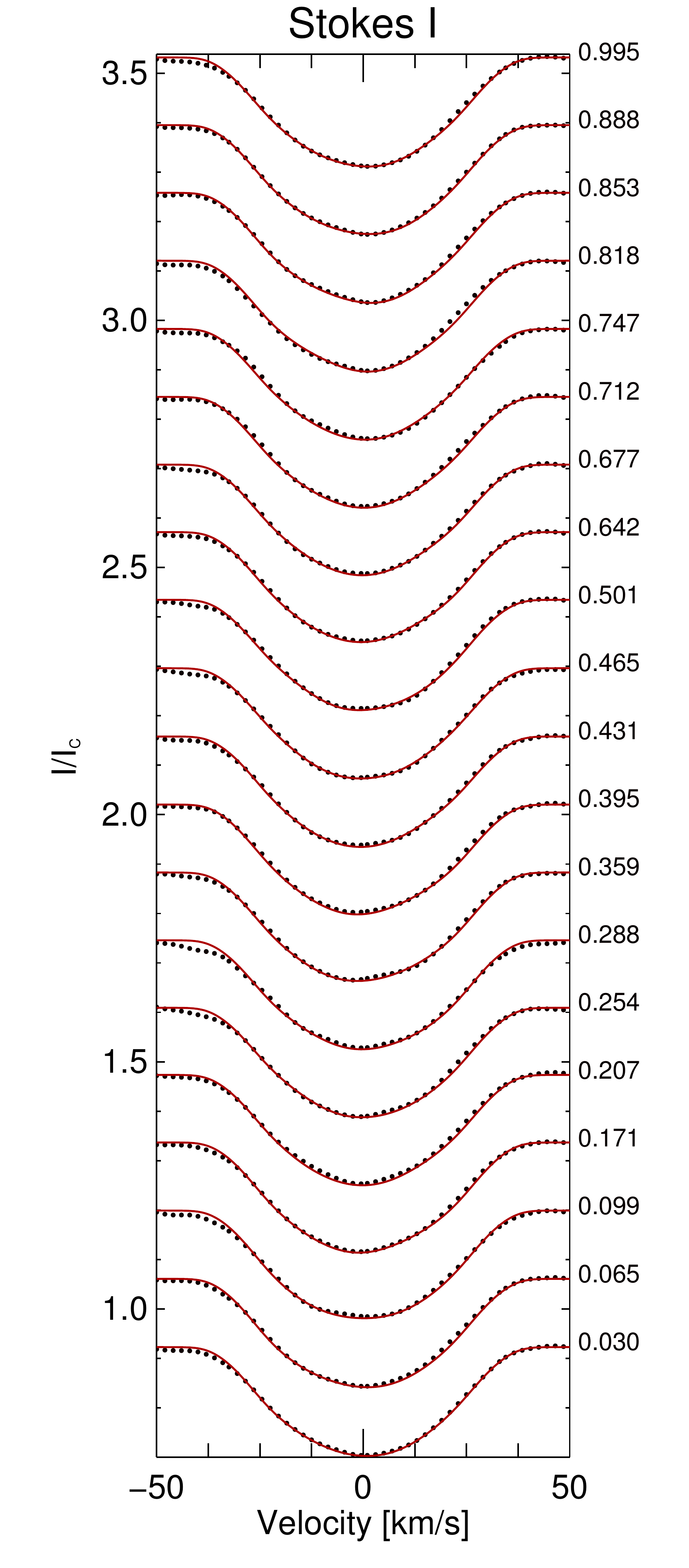}

\centering
\vspace{0.35cm}
\hspace{-2.1cm}\large{S23}\hspace{5.47cm}\large{S24}\hspace{5.47cm}\large{S25}
\vspace{-0.35cm}

\includegraphics[width=0.565\columnwidth]{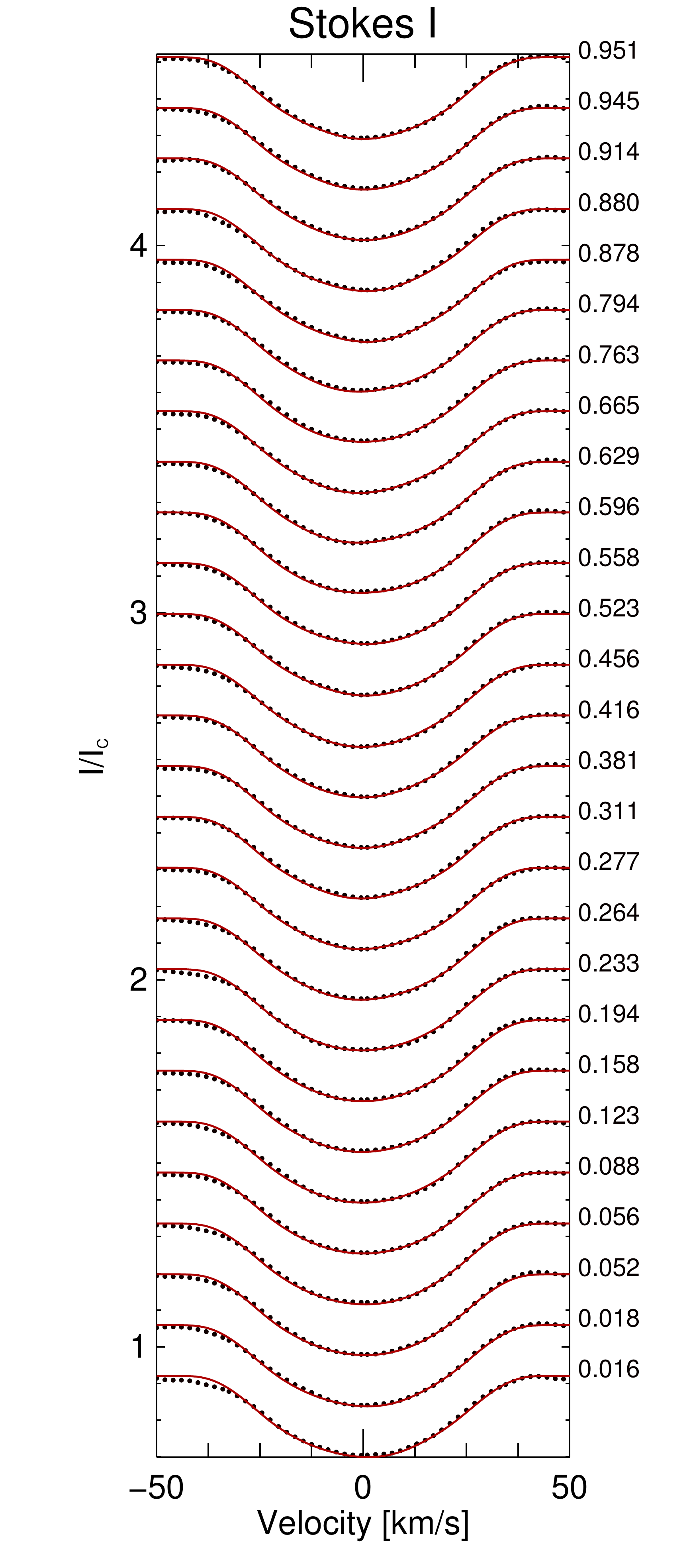}\hspace{9mm}
\includegraphics[width=0.565\columnwidth]{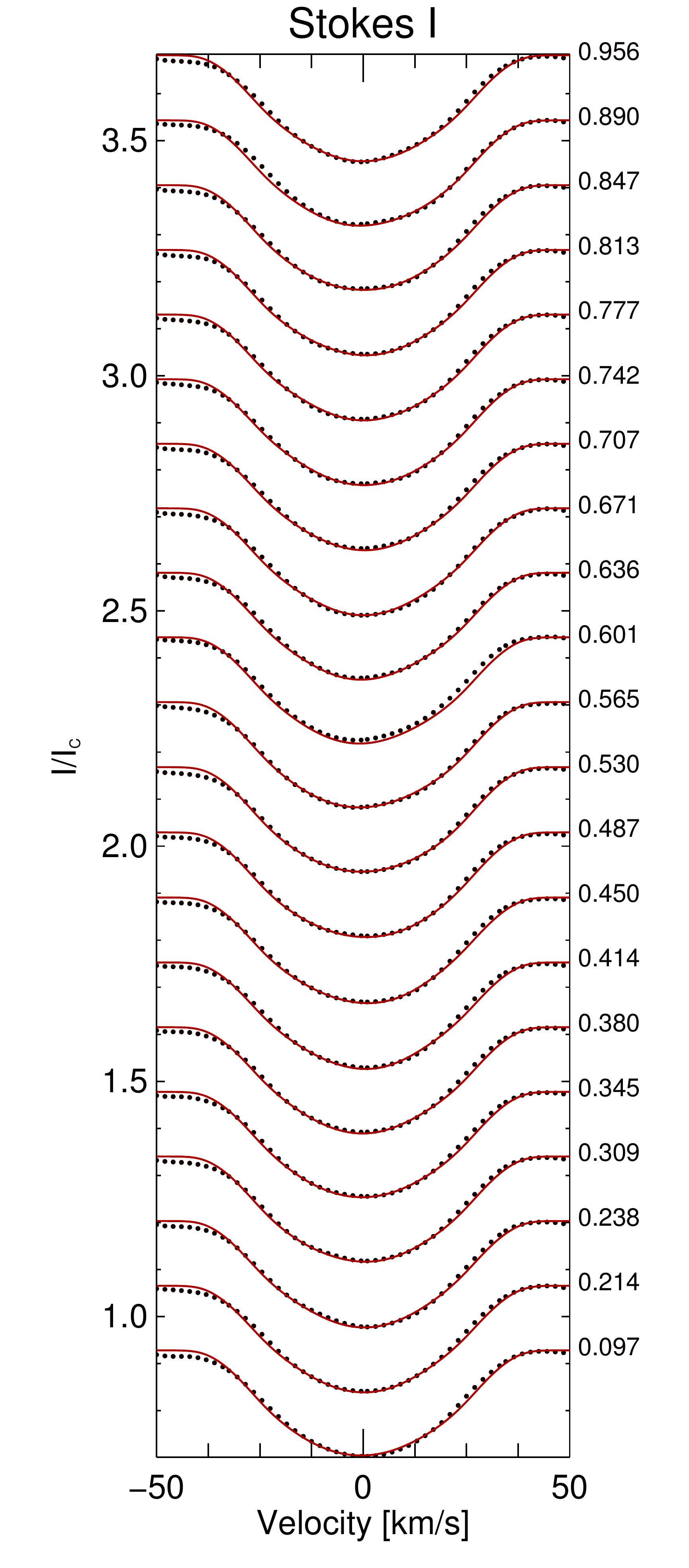}\hspace{9mm}
\includegraphics[width=0.565\columnwidth]{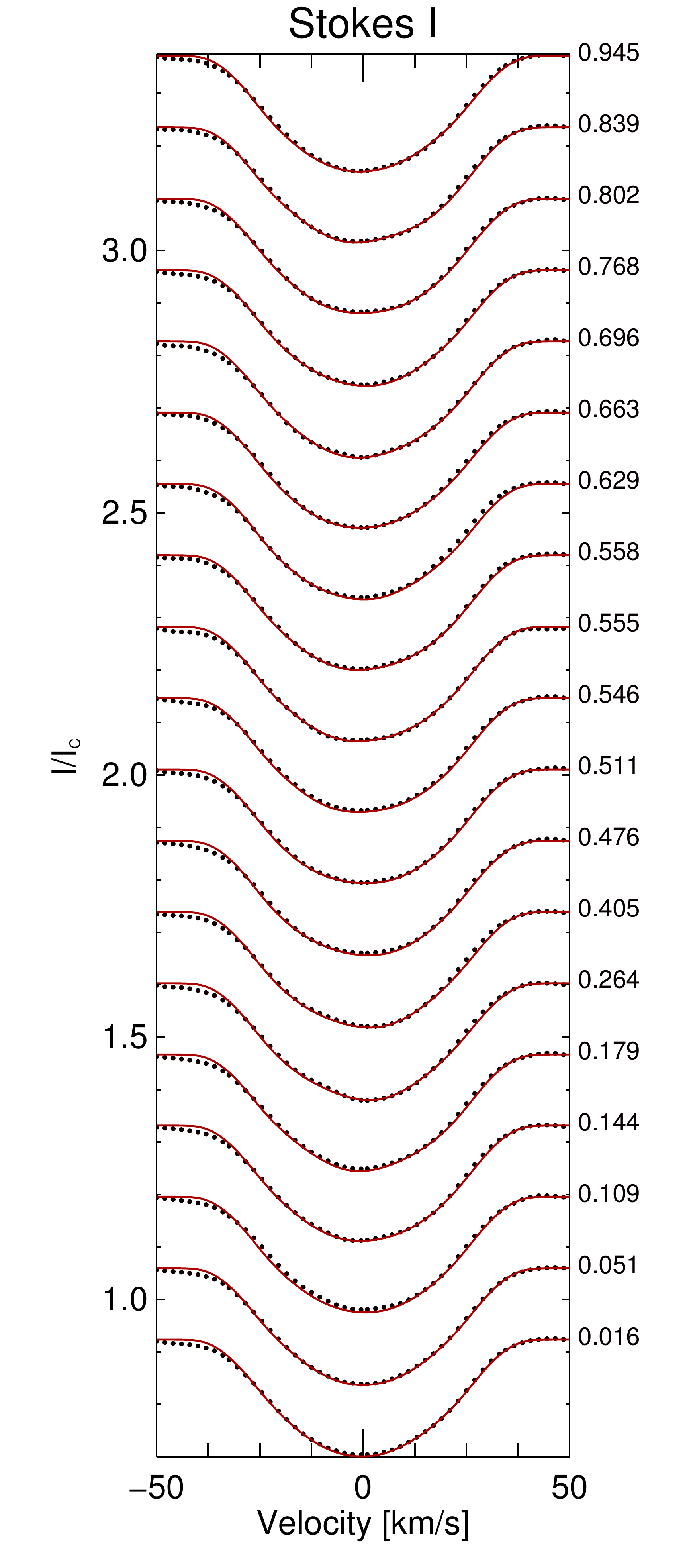}
\caption{Line profile fits for the Doppler reconstructions shown in Figs.~\ref{dis11}--\ref{dis2x}.
The phases of the individual observations are listed on the right side of the panels.}
\label{proffits1}
\end{figure*}


\begin{figure*}[tb]
\centering
\vspace{0.35cm}
\hspace{-2.1cm}\large{S31}\hspace{5.47cm}\large{S32}
\vspace{-0.35cm}

\includegraphics[width=0.565\columnwidth]{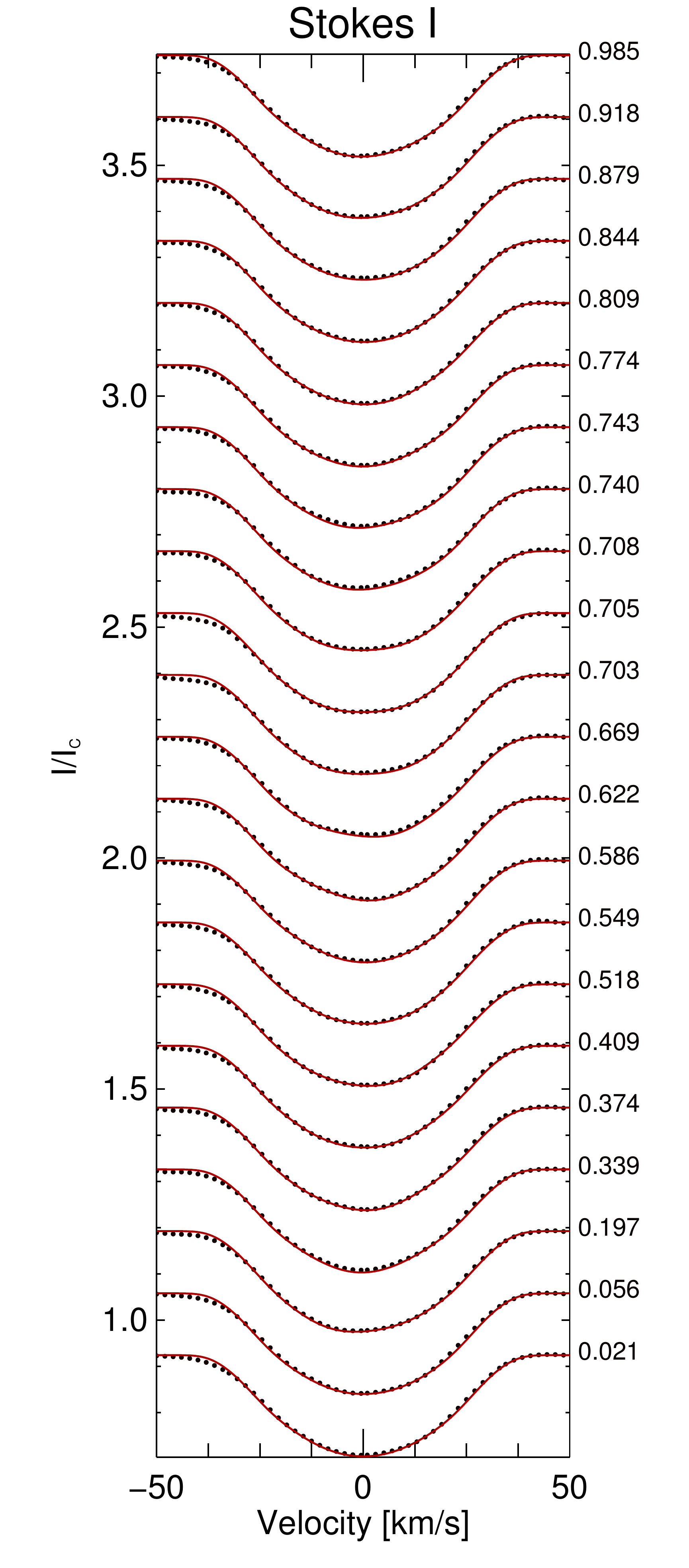}\hspace{9mm}
\includegraphics[width=0.565\columnwidth]{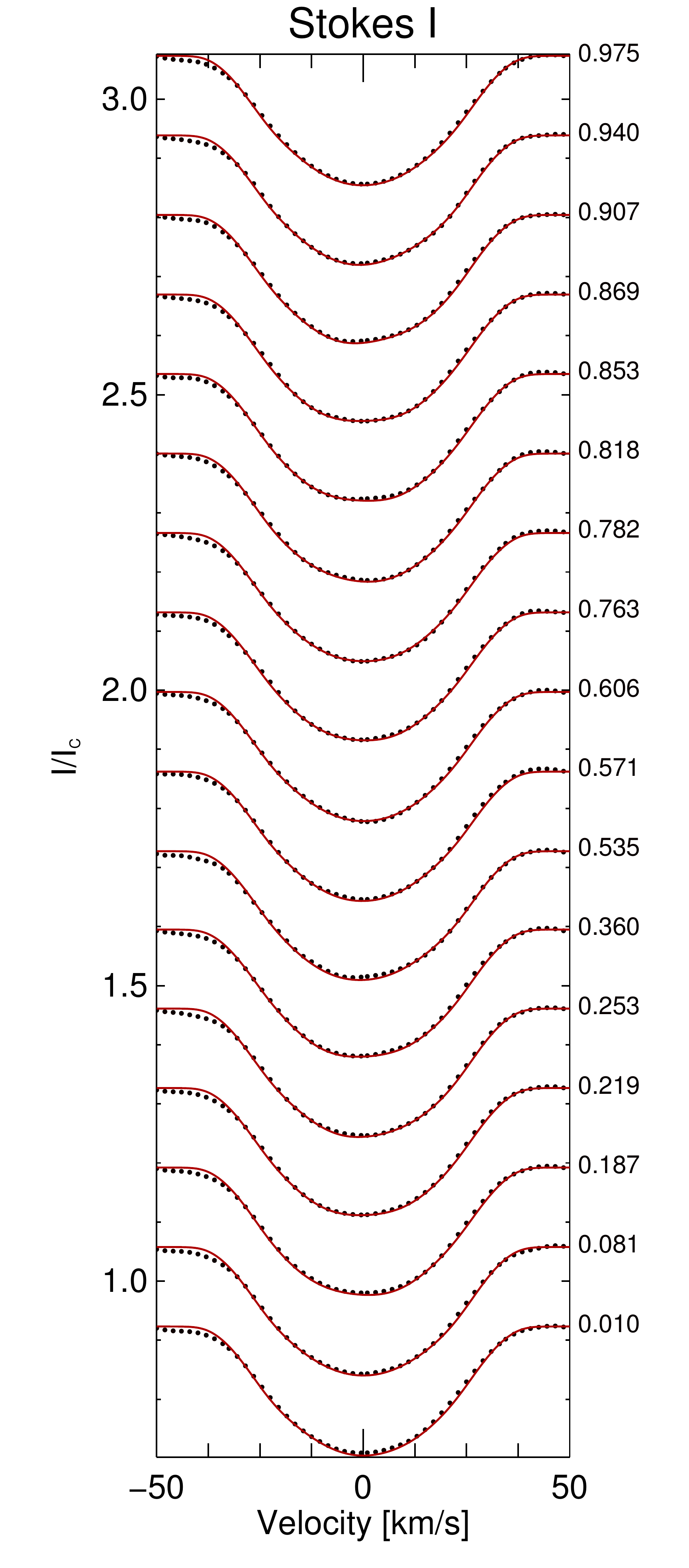}

\centering
\vspace{0.35cm}
\hspace{-2.1cm}\large{S41}\hspace{5.47cm}\large{S51}\hspace{5.47cm}\large{S61}
\vspace{-0.35cm}

\includegraphics[width=0.565\columnwidth]{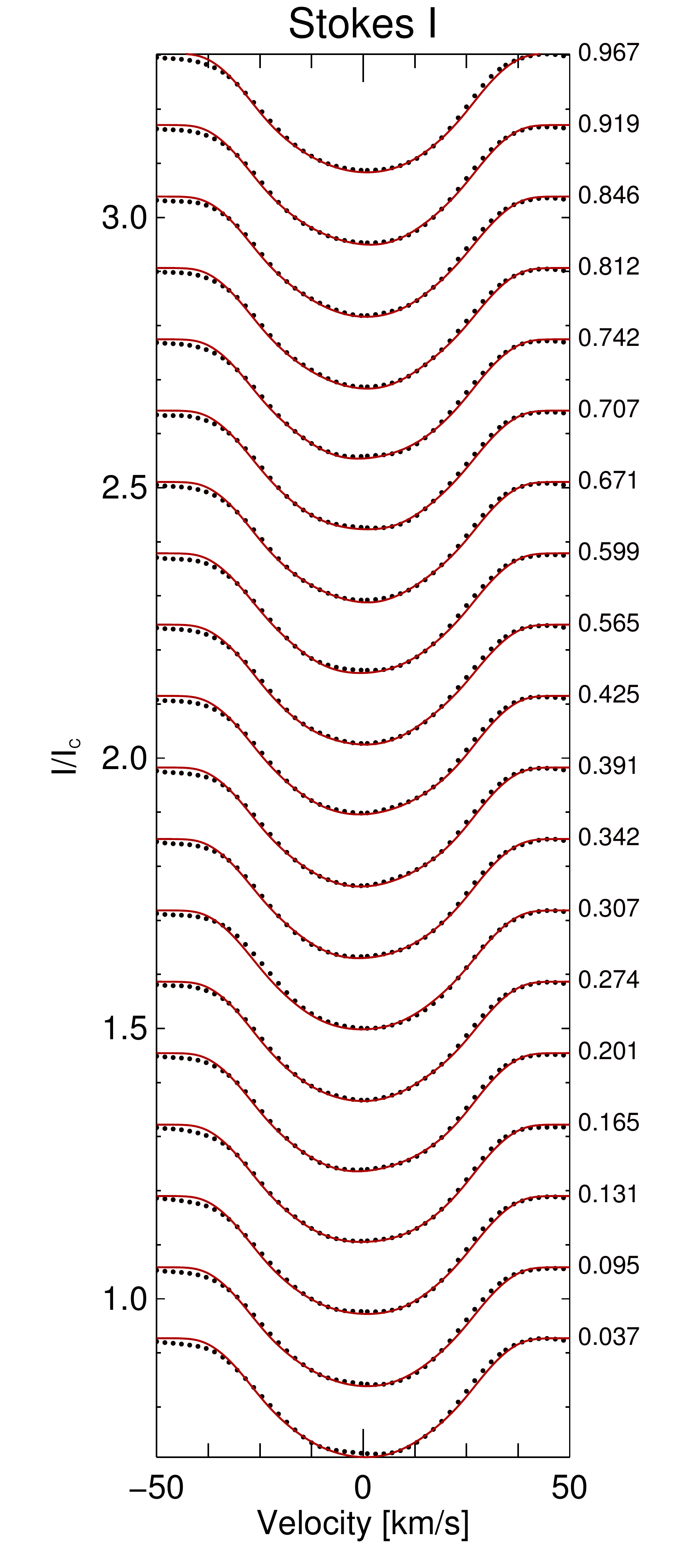}\hspace{9mm}
\includegraphics[width=0.565\columnwidth]{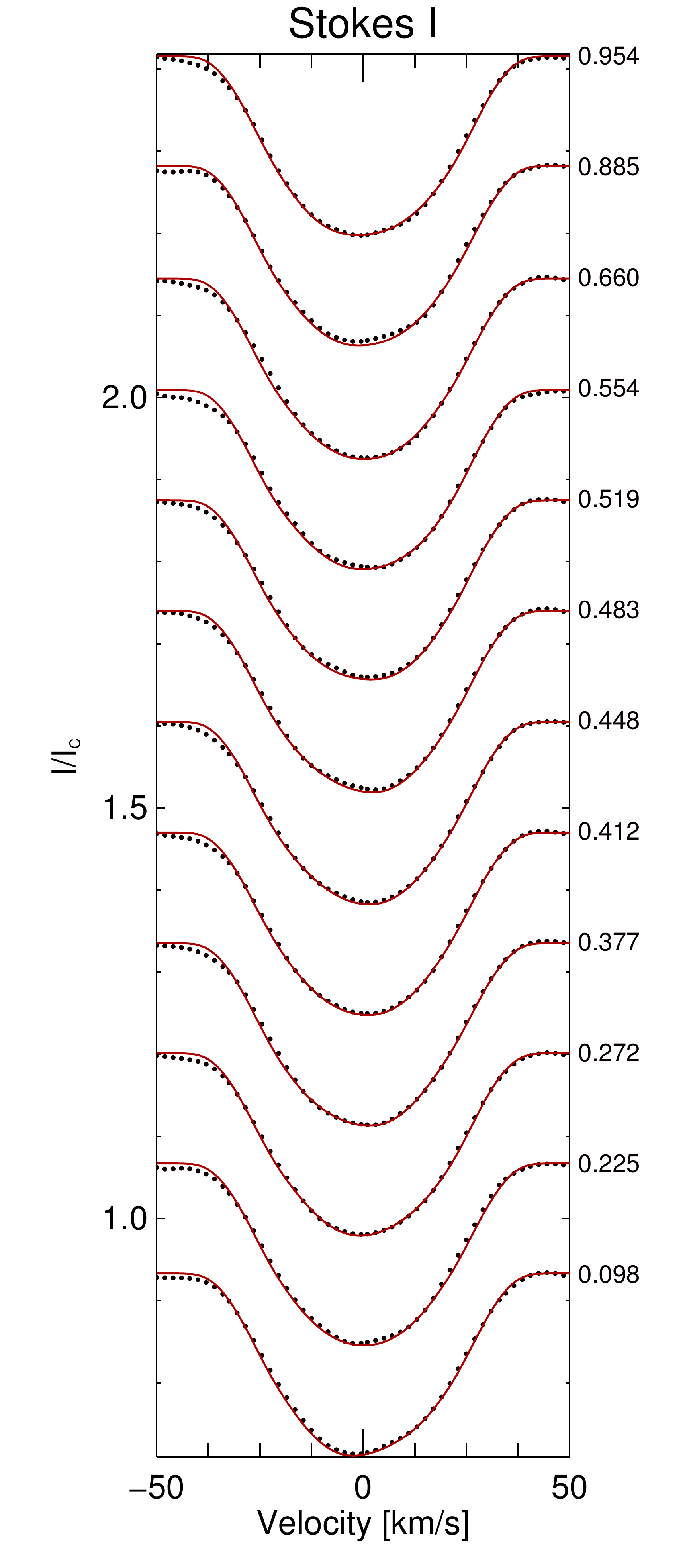}\hspace{9mm}
\includegraphics[width=0.565\columnwidth]{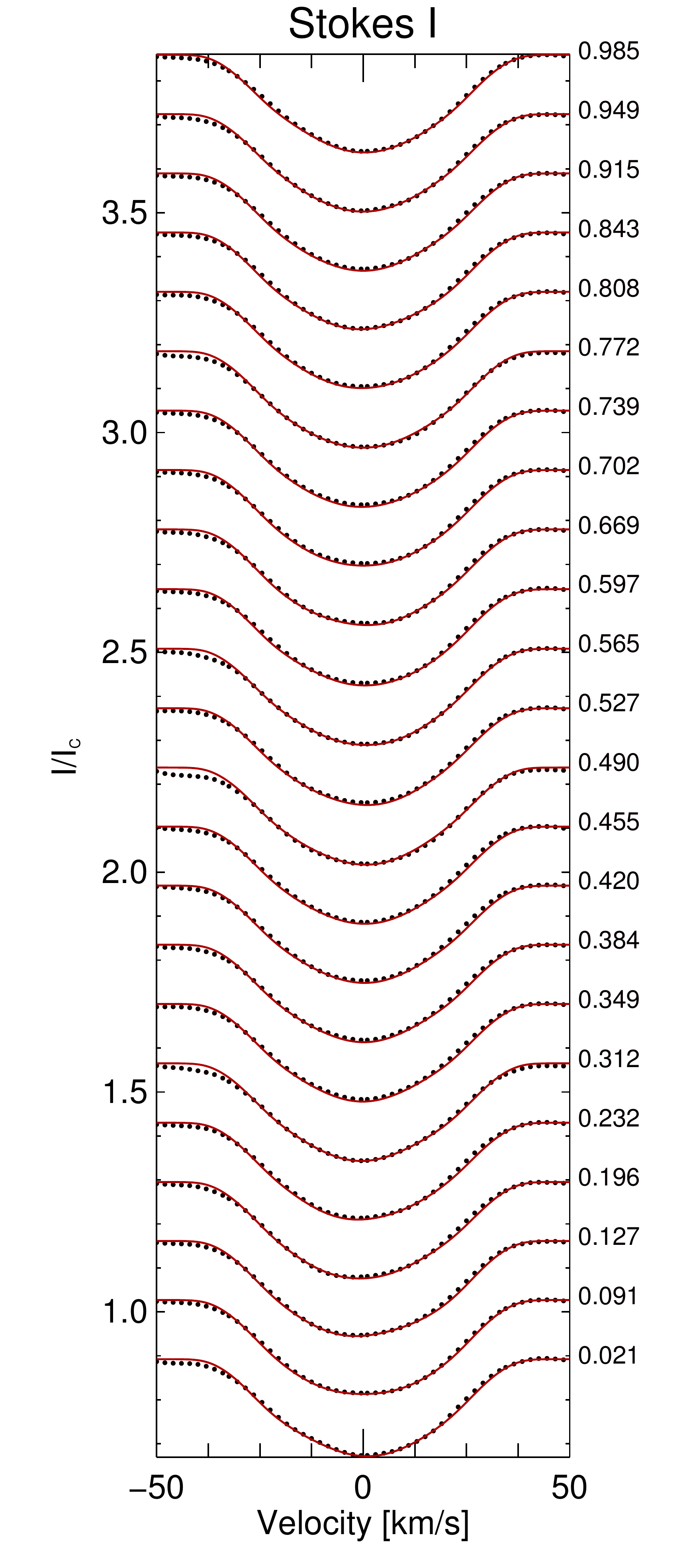}
\caption{Line profile fits for the Doppler reconstructions shown in Figs.~\ref{dis3x}--\ref{dis61}.
The phases of the individual observations are listed on the right side of the panels.}
\label{proffits2}
\end{figure*}

\end{appendix}

\appendix
\end{document}